\documentclass[12pt,twoside,english]{article}
\usepackage[T1]{fontenc}
\usepackage[latin1]{inputenc}
\usepackage{geometry}
\usepackage{babel}
\usepackage{graphics}

\begin{document}
\begin{titlepage}
\setcounter{page}{1}
\rightline{}
\vfill
\begin{center}
{\Large \bf Deeply Virtual Neutrino Scattering (DVNS)}
\vfill

\vfill

{\large $^{a}$Paolo Amore, $^{b,c}$Claudio Corian\`{o} and  $^{b}$Marco Guzzi}\\ 
\vspace{.12in}
{\it $^{a}$Facultad de Ciencias, Universidad de Colima, \\
Av. 25 de Julio 965, Colima Colima,28045 Mexico}\\
\vspace{.12in}
{\it  $^{b}$Dipartimento di Fisica, Universit\`{a} di Lecce \\
and INFN Sezione di Lecce \\ Via Arnesano 73100 Lecce, Italy}\\
\vspace{.25cm}

{\it $^{c}$ Department of Physics and Institute of Plasma Physics\\ 
University of Crete and FORTH\\
Heraklion, Greece}

\end{center}
\vfill

\begin{abstract}

We introduce the study of neutrino scattering off protons 
in the deeply virtual kinematics, which describes 
under a unified formalism elastic and deep inelastic neutrino scattering. 
A real final state photon and a recoiling nucleon are detected in the few GeV  ($|t|\sim 0.2-5$ GeV)  region of momentum transfer. 
This is performed via an extension of the notion of deeply virtual Compton scattering, or DVCS, to the case of a neutral current exchange. 
The relevance of this process and of other similar exclusive processes 
for the study of neutrino interactions in neutrino factories for GeV 
neutrinos is pointed out. 

\end{abstract}
\smallskip
\end{titlepage}
\setcounter{footnote}{0}


\def\beq{\begin{equation}}
\def\eeq{\end{equation}}
\def\beqn{\begin{eqnarray}}
\def\eeqn{\end{eqnarray}}
\def\ba{\begin{eqnarray}}
\def\ea{\end{eqnarray}}
\def\ie{{\it i.e.}}
\def\eg{{\it e.g.}}
\def\half{{\textstyle{1\over 2}}}
\def\nicefrac#1#2{\hbox{${#1\over #2}$}}
\def\third{{\textstyle {1\over3}}}
\def\quarter{{\textstyle {1\over4}}}
\def\m{{\tt -}}
\def\p{{\tt +}}
\def\slash#1{#1\hskip-6pt/\hskip6pt}
\def\slk{\slash{k}}
\def\GeV{\,{\rm GeV}}
\def\TeV{\,{\rm TeV}}
\def\y{\,{\rm y}}
\def\ds{\slash}
\def\l{\langle}
\def\r{\rangle}
\def\xprime{x^{\prime}}
\def\xprimetwo{x^{\prime\prime}}
\def\zprime{z^{\prime}}
\def\xprimbar{\overline{x}^\prime}
\def\xprim2bar{\overline{x}^{\prime\prime}}
\def\ptbold{\mbox{\boldmath$p$}_T}
\def\ktbold{\mbox{\boldmath$k$}_T}
\def\ktboldbar{\mbox{\boldmath$\overline{k}$}_T}
\def\beq{\begin{equation}}
\def\eeq{\end{equation}}
\def\tr{{\bf tr}}
\def\P{P^\mu}
\def\Pb{\overline{P}^\mu}
\def\BOX#1#2#3#4#5{\hskip#1mm\raisebox{#2mm}[#3mm][#4mm]{$#5$}}
\def\VBOX#1#2{\vbox{\hbox{#1}\hbox{#2}}}

\setcounter{footnote}{0}
\newcommand{\beqa}{\begin{eqnarray}}
\newcommand{\eeqa}{\end{eqnarray}}
\newcommand{\eps}{\epsilon}

\pagestyle{plain}
\setcounter{page}{1}

\section{Introduction} 
There is no doubt that the study of neutrino masses and of flavour mixing 
in the leptonic sector will play a crucial role for uncovering new physics beyond the Standard Model 
and to test Unification. In fact, the recent discovery of neutrino oscillations 
in atmospheric and solar neutrinos (see \cite{Valle} for an overview) has raised 
the puzzle of the origin of the mass hierarchy among the various neutrino flavours, 
a mystery which, at the moment, remains unsolved. 
The study of the mixing among the leptons also raises the possibility of detecting 
possible sources of CP violation in this sector as well.
It seems then obvious that the study of these aspects of flavour physics requires the exploration of a new energy range for neutrino production and detection beyond the one which is accessible at this time. 

For this purpose, 
several proposals have been presented recently for neutrino factories, where a beam, primarily made of muon neutrinos produced 
at an accelerator facility, is directed 
to a large volume located several hundreds kilometers away at a second facility. The goal 
of these experimental efforts is to uncover various 
possible patterns of mixings among flavours - using the large distance 
between the points at which neutrinos are produced and detected - 
in order to study in a more detailed and ``artificial'' way the phenomenon of oscillations. 
Detecting neutrinos at this higher energies is an aspect that deserves special attention since several 
of these experimental proposals \cite{minos,Marciano1,Mangano} require a nominal energy of the neutrino beam in the few GeV region.  
We recall that the incoming neutrino beam, scattering off deuteron or other heavier targets at the detector facility, 
has an energy which covers, in the various proposals, both the resonant, the quasi-elastic (in the GeV range) and the deep inelastic region (DIS) at higher energy.   
In the past, neutrino scattering on nucleons has been observed over a wide interval of energy, ranging from few MeV up to 100 GeV, and these studies have been of significant help for uncovering the structure of the fundamental interactions in the Standard Model. Generally, one envisions contributions to the scattering cross section either in the low energy region, such as in neutrino-nucleon elastic scattering, or in the deep inelastic scattering (DIS) region. Recent developments in perturbative QCD have emphasized that exclusive 
(see \cite{Sterman}) and inclusive processes can be unified under a general treatment using a factorization approach in a generalized kinematical domain. The study of this domain, termed deeply virtual Compton scattering, or DVCS, is an area of investigation of wide theoretical interest, with experiments planned in the next few years at JLAB and at DESY. The key constructs of the DVCS domain are the non-forward parton distributions, where the term {\em non-forward} is there to indicate the asymmetry between the initial and final state typical of a true Compton process, in this case appearing not through unitarity, such as in DIS, but at amplitude level. 

In this work, after a brief summary of the generalities of process, we discuss its generalization to the case of neutral currents. We recall that DVCS has been extensively studied in the last few years for electromagnetic interactions. The extension of DVCS to the case of neutral currents is presented here, 
while the charged current version will be presented in a forthcoming paper. 
Also, in this work we will just focus on the lowest order contributions to the process, named by us Deeply Virtual Neutrino Scattering (DVNS) in order to distinguish 
it from standard DVCS, while a renormalization group analysis of the factorized amplitude, which requires an inclusion of the modifications induced by the evolution will also be presented elsewhere. 
The application of the formalism that we develop here also needs 
a separate study of the isoscalar cross sections together 
with a detailed analysis of the various experimental constraints at neutrino factories 
in order to be applicable at forthcoming experiments. 

\section{The Generalized Bjorken Region and DVCS}

Compton scattering has been investigated in the near past by several groups, since the original work by Ji and Radyushkin \cite {Ji1, Rad3, Rad1}.
Previous work on the generalized Bjorken region, which includes DVCS and predates the ``DVCS period'' can be found in \cite{Geyer}.

\begin{figure}[t]
{\par\centering \resizebox*{12cm}{!}{\includegraphics{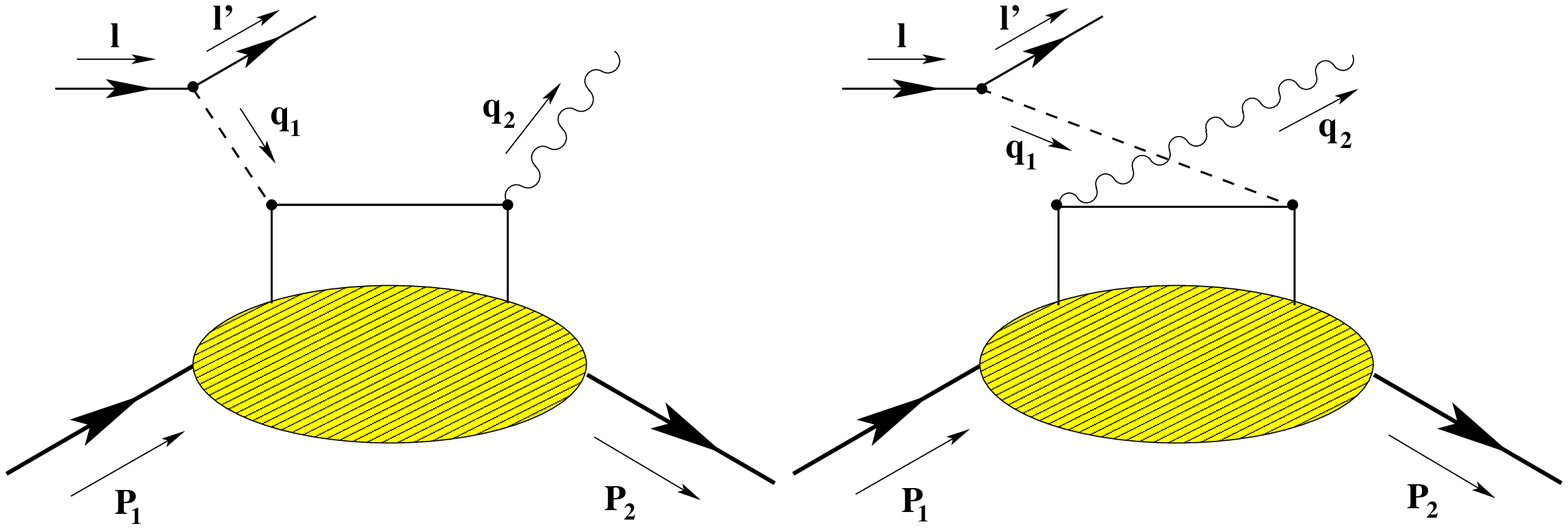}} \par}
\caption{Leading hand-bag diagrams for the process}
\label{DVCS_1.eps}
\end{figure}

A pictorial description of the process we are going to illustrate is given in Fig~\ref{DVCS_1.eps} where a neutrino of momentum $l$ scatters off a nucleon of momentum $P_1$ by an interaction with a neutral current; from the final state a photon and a nucleon emerge, of momenta $q_2$ and $P_2$ respectively, while the momenta of the final lepton is $l'$. 
The process is described in terms of new constructs of the parton model termed 
generalized parton distributions (GPD) or also non-forward (off-forward) parton distributions.
We recall that the regime for the study of GPD's is characterized by a deep virtuality of the exchanged photon in the initial interaction ($e +p\to e +p +\gamma$) ( $ Q^2 \approx $  2 GeV$^2$), with the final state photon kept on-shell; large energy of the hadronic system ($W^2 > 6$ GeV$^2$) above the resonance domain and small momentum transfers $|t| < 1$ GeV$^2$. 
The process suffers of a severe Bethe-Heitler (BH) background, with photon emission taking place from the lepton. Therefore, in the relevant region, characterized by large $Q^2$ and small $t$, the dominant Bethe-Heitler background $(\sim 1/t)$ and the $1/Q$ behaviour of the DVCS scattering amplitude render the analysis quite complex. 
From the experimental viewpoint a dedicated study of the interference BH-VCS is required in order to explore the generalized Bjorken region, and this is done by measuring asymmetries. 
Opting for a symmetric choice for the defining momenta, we use as independent variables the average of the hadron and gauge bosons momenta 

\beq
P_{1,2}= \bar{P} \pm \frac{\Delta}{2}\,\,\,\,\,\,\,\, q_{1,2}= \bar{q} \mp \frac{\Delta}{2}
\eeq

with $-\Delta=P_2-P_1$ being the momentum transfer. Clearly 

\beq
\bar{P}\cdot \Delta=0,\,\,\,\,\, t=\Delta^2 \,\,\,\,\,\,\, \bar{P}^2=M^2 - \frac{t}{4}
\eeq

and $M$ is the nucleon mass. There are two scaling variables which are identified in the process, since 3 scalar products  can grow large in the generalized Bjorken limit: $\bar{q}^2$, $\Delta\cdot q$, $\bar{P}\cdot \bar{q}$. 

The momentum transfer $t=\Delta^2$ is a small parameter in the process. Momentum asymmetries between the initial and the final state nucleon are measured by two scaling parameters, $\xi$ and $\eta$, related to ratios of the former invariants 

\beq
\xi=-{\bar{q}^2\over 2 \bar{P}\cdot {\bar q}} \,\,\,\,\,\,\,\,\,\,\,\, \eta={\Delta\cdot \bar{q}\over 2 \bar{P}\cdot \bar{q}}
\eeq

where $\xi$ is a variable of Bjorken type, expressed in terms of average momenta rather than nucleon and Z-boson momenta.  
The standard Bjorken variable $x= - q_1^2/( 2 P_1\cdot q_1)$ is trivially related to $\xi$ in the $t=0$ limit.  
In the DIS limit $(P_1=P_2)$ $\eta=0$ and $x=\xi$, while in the DVCS limit $\eta=\xi$ and $x=2\xi/(1 +\xi)$, as one can easily deduce from the relations 

\beq
q_1^2=\left(1 +\frac{\eta}{\xi}\right) \bar{q}^2 +\frac{t}{4},\,\,\,\,\,\,\,\,\,\,
q_2^2=\left(1 -\frac{\eta}{\xi}\right) \bar{q}^2 +\frac{t}{4}.
\eeq

We introduce also the inelasticity parameter $y=P_1\cdot l/(P_1\cdot q_1)$ which measures the fraction of the total energy that is transferred to the final state photon. Notice also that $\xi= \frac{\Delta^+}{2 \bar{P}^+}$ measures the ratio between the plus component of the momentum transfer and the average momentum. 
A second scaling variable, related to $\xi$ is $\zeta={\Delta^+}/{{P_1}^+}$, which coincides with Bjorken $x$ $(x=\zeta)$ when $t=0$. 

$\xi$, therefore, parametrizes the large component of the momentum transfer $\Delta$, which can be generically described as 

\beq
\Delta= 2 \xi \bar{P} + \hat{\Delta}
\eeq

where all the components of $\hat{\Delta}$ are $O(\sqrt{t})$ \cite{Rad&Wei}. 

\section{DIS versus DVNS}

\begin{figure}[t]
{\par\centering \resizebox*{6cm}{!}{\includegraphics{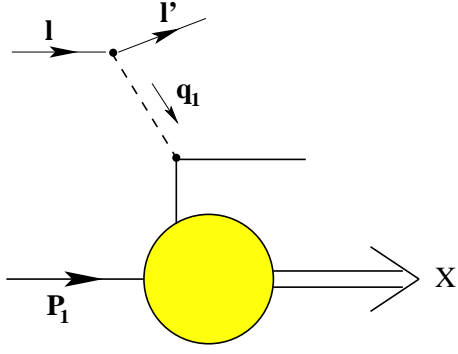}} \par}
\caption{Leading diagram for a generic DIS process} 
\label{DIS}
\end{figure}

In the study of ordinary DIS scattering of neutrinos on nucleons (see Fig.~\ref{DIS}), the relevant current correlator is obtained from the T-product of two neutral currents acting on a forward nucleon state of momentum $P_1$ 

\ba
j_Z^\mu &\equiv& \overline{u}(l') \gamma^\mu \left( -1 + 4 \sin^{2}\theta_W + \gamma_5 \right) u(l)
\ea
where $\theta_W$ is the Weinberg angle, $l$ and $l'$ are the initial and final-state lepton. The relevant correlator is given by
\beq
T_{\mu\nu}(q_1^2,\nu)=i \int d^4 z e^{i q\cdot z}\langle P_1|T(J_Z^\mu(\xi)J_Z^{\nu}(0)
|P_1\rangle 
\eeq
with $\nu=E- E'$ being the energy transfered to the nucleon, $P_1$ is the initial-state nucleon 4-momenta and $q_1 = l - l'$ is the momentum transfered.
The hadronic tensor $W_{\mu\nu}$ is related to the imaginary part of this correlator by the optical theorem. We recall that for an inclusive electroweak process mediated by neutral currents the hadronic tensor (for unpolarized scatterings) is identified in terms of 3 independent structure functions at leading twist

\beq
W_{\mu\nu}=\left(- g_{\mu\nu}+ {q_{1\mu}q_{1\nu}\over q_1^2}\right) W_1(\nu,Q^2)+ 
{\hat{P_1}^\mu\hat{P_1}^\nu\over P_{1}^2}{W_2(\nu,Q^2)\over M^2} 
-i\epsilon_{\mu\nu\lambda\sigma}{q_1^\lambda P_1^\sigma}{W_3(\nu,Q^2)\over 2 M^2}
\eeq

where transversality of the current is obvious since $\hat{P_1}^\mu= P_1^\mu - q_1^\mu{P_1\cdot q_1}/{q_1^2}$. 

The analysis at higher twists is far more involved and the total number of structure functions appearing is 14 if we include polarization effects. These are fixed by the requirements of Lorenz covariance and time reversal invariance, neglecting small CP-violating effects from the CKM matrix. 
Their number can be reduced to 8 if current conservation is imposed, which is equivalent to requiring that contributions proportional to non-vanishing current quark masses can be dropped (see also \cite{Blum}).
We recall that the DIS limit is performed by the identifications 

\ba
M W_1(Q^2,\nu)&=&F_1(x,Q^2) \nonumber \\
\nu W_2(Q^2,\nu)&=&F_2(x,Q^2)\nonumber \\
\nu W_3(Q^2,\nu)&=&F_3(x,Q^2), 
\ea

in terms of the standard structure functions $F_1$, $F_2$ and $F_3$.
There are various ways to express the neutrino-nucleon DIS cross section, either in terms of $Q^2$ and the energy transfer, in which the scattering angle $\theta$ is integrated over, or as a triple cross section in $(Q^2,\nu,\theta)$, or yet in terms of the Bjorken variable $x$, inelasticity $y$ and the scattering angle $(x,y,\theta)$. This last case is close to the kinematical setup of our study. In this case the differential Born cross section in DIS is given by 

\beq
\label{e1}
\frac{d^3 \sigma}{dx dy d\theta} = \frac{y \alpha^2}
{Q^4}\sum_{i}\eta_{i}(Q^2) L_{i}^{\mu\nu} W_{i}^{\mu\nu},
\eeq

the index $i$ denotes the different current contributions, ( $i = |\gamma|^2, |\gamma Z|, |Z|^2$ for the neutral current) and $\alpha$ denotes the fine structure constant. By $\theta$ we indicate the azimuthal angle of the final-state lepton, while $y =  (P_1\cdot q_1)/(l\cdot P_1)$ is the inelasticity parameter, and $Q^2 = -q_1^2$.
The factors $\eta_i(Q^2)$ denote the ratios of the corresponding propagator terms to the photon propagator squared,

\ba
\label{etas}
\eta^{|\gamma|^2}(Q^2)  &=& 1 ,\nonumber\\
\eta^{|\gamma Z|}(Q^2)  &=& \frac{G_F M_Z^2}{2\sqrt{2}\pi\alpha}\frac{Q^2}{Q^2+M_Z^2},\nonumber\\
\eta^{|Z|^2}(Q^2)       &=& (\eta^{|\gamma Z|})^2(Q^2).\nonumber\\
\ea

where $G_F$ is the Fermi constant and $M_Z$ is the mass of the $Z$ boson while the leptonic tensor has the form

\beq
\label{e2}
L_{\mu\nu}^i=\sum_{\lambda^\prime}\left[\bar
u(k^\prime,\lambda^\prime)\gamma_\mu(g_V^{i_1}+g_A^{i_1}
\gamma_5)u(k,\lambda)\right]^\ast\bar
u(k^\prime,\lambda^\prime)\gamma_\nu(g_V^{i_2}
+g_A^{i_2}\gamma_5)u(k,\lambda).
\eeq
In the expression above $\lambda$ and $\lambda^{\prime}$, denote the initial and final-state helicity 
of the leptons. The indices $i_1$ and $i_2$ refer to the sum appearing in eq. (\ref{e1})   

\begin{equation}
\label{ac}
\begin{array}{lcllcl}
g_V^{\gamma} &=& 1, & g_A^{\gamma} &=& 0, \\
g_V^Z &=& -\frac{1}{2}+2 \sin^2\theta_W,& g_A^Z &=& \frac{1}{2}, \\
\end{array}
\end{equation}

In the case of neutrino/nucleon interaction mediated by the neutral current the dominant diagram for this process appears in Fig.~\ref{DIS}. A similar diagram, with the obvious modifications, describes also charged current exchanges.  
We recall that the unpolarized cross section is expressed in terms of $F_1$ and $F_2$, since $F_3$ disappears in this special case, 
and in particular, after an integration over the scattering angle of the final state neutrino one obtains                             

\beq
\frac{d^2\sigma}{dx dy}  = 2 \pi S {\alpha^2\over Q^4} (g_V^Z)^2 \eta^{|\gamma Z|}(Q^2)
\left( 2 x y^2 F_1 + 2(1-x- {xy M^2\over S})F_2\right). 
\eeq

where $S=2 M E_\nu$ is the nucleon-neutrino center of mass energy.

We also recall that in this case the cross section in the parton model is given by 

\ba
\frac{d^2\sigma}{dx dy}=\frac{G_{F}^2 M E_{\nu}}{2\pi}\left( \frac{M_{Z}^2}{Q^2+M_Z^2}\right)^2\left[x q^{0}(x,Q^2)+x\bar{q}^{0}(x,Q^2)(1-y)^2\right]
\ea
where $q^{0}(x,Q^2)$ and $\bar{q}^{0}(x,Q^2)$ are linear combinations of parton distributions 

\ba
q^{0}(x,Q^2)&=&\left[\frac{u_{v}(x,Q^2)+d_{v}(x,Q^2)}{2}+\frac{\bar{u}(x,Q^2)+\bar{d}(x,Q^2)}{2}\right] \left(L_u^2 + L_d^2\right)\nonumber\\
&+&\left[\frac{\bar{u}(x,Q^2)+\bar{d}(x,Q^2)}{2}\right]\left(R_u^2 + R_d^2\right)\nonumber\\
\bar{q}^{0}(x,Q^2)&=&\left[\frac{u_v(x,Q^2)+d_v(x,Q^2)}{2}+\frac{\bar{u}(x,Q^2)+\bar{d}(x,Q^2)}{2}\right]\left(R_u^2+R_d^2\right)\nonumber\\
&+&\left[\frac{\bar{u}(x,Q^2)+\bar{d}(x,Q^2)}{2}\right]\left(L_u^2+L_d^2\right)\nonumber\\
\ea
with

\ba
&&L_u =1-\frac{4}{3}\sin^2\theta_{W}\,,\hspace{1.5 cm}L_d =-1+\frac{2}{3}\sin^2\theta_{W}\nonumber\\
&&R_u =-\frac{4}{3}\sin^2\theta_{W}\,,\hspace{1.8 cm}R_d =\frac{2}{3}\sin^2\theta_{W}\nonumber\\
\ea
and we have identified the sea contributions $u_{s}$ and $d_{s}$ with $\bar{u}$ and $\bar{d}$ rispectively.

\begin{figure}[t]
{\par\centering \resizebox*{10cm}{!}{\rotatebox{-90}{\includegraphics{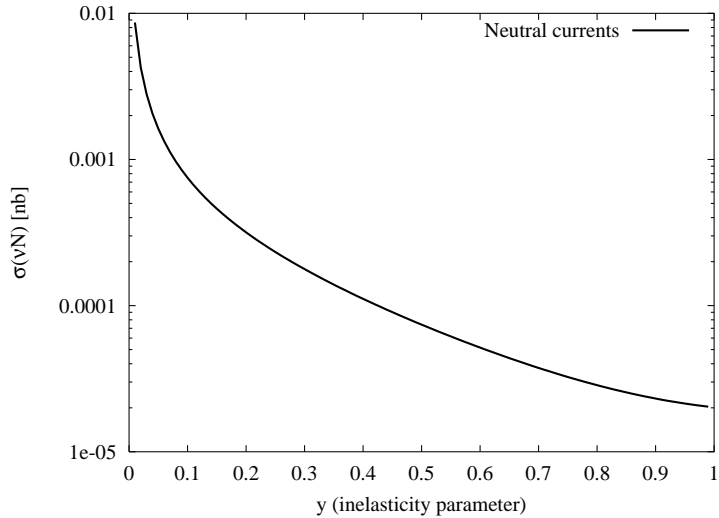}}}\par}
\caption{The cross section of a neutrino process of DIS-type 
at $x\approx 0.1$ with neutral current at ultrahigh energy}
\label{sigma_Dis.ps}
\end{figure}

Let's now move to the nonforward case. Here, when a real photon is present in the final state, the relevant correlator is given by 

\beq
T_{\mu\nu}(q_1^2,\nu)=i \int d^4 z e^{i q\cdot z}\langle\bar{ P} - {\Delta\over 2}|T(J_Z^\mu(-z/2)J_{\gamma}^{\nu}(z/2)|\bar{P} +{\Delta\over 2}\rangle. 
\eeq

The dominant diagrams for this process appears in Fig~\ref{DVCS_1.eps}.
We impose Ward identities on both indices, which is equivalent to requiring that terms proportional to the quark masses in $\partial\cdot J_Z$ are neglected. This approximation is analogous 
to the one performed in the forward case in order to reduce the structure functions from 8 to 3 
(in the absence of any polarization), imposing symmetric trasversality conditions on the weak currents 

\beq 
\left(\bar{P}^\mu -{\Delta^\mu\over 2}\right) T_{\mu\nu}=0\,\,\,\,\,\,\,\,\,
\left(\bar{P}^\mu +{\Delta^\mu\over 2}\right) T_{\mu\nu}=0.
\eeq

The leading twist contribution to DVNS is obtained by performing a collinear expansion of the loop momentum 
of the hand-bag diagram and neglecting terms of order $O(\Delta_\perp^2/Q^2)$ and $M^2/Q^2$. Transversality is satisfied at this order.
 Violation of transversality condition in the hand-bag approximation is analogous to the DVCS case, 
where it has been pointed out that one has to include systematically ``kinematical'' twist-3 operators,  which 
appear as total derivatives of twist-2 operators \cite{Rad&Wei} \cite{Penttinen} in order to restore it.  

For the parametrization of the hand-bag diagram (Fig.~\ref{DVCS_1.eps}) 
we use the light-cone decomposition in terms of 2 four-vectors $(n,\tilde{n})$, where 

\begin{eqnarray}
\tilde{n}^\mu &=& \Lambda (1,0,0,1) \nonumber \\
n^\mu &=& \frac{1}{2 \Lambda} (1,0,0,-1) \nonumber \\
\tilde{n}^2 &=& n^2 = 0 \ \ \ , \ \ \ \tilde{n}\cdot n = 1. \nonumber 
\end{eqnarray}

At the same time we set 

\begin{eqnarray}
P_1^\mu&=& (1 +\xi)\tilde{n}^{\mu} +(1-\xi){\overline{M}\over 2}n^\mu - {\Delta_\perp^\mu\over 2} \nonumber \\
P_2^\mu&=& (1 +\xi)\tilde{n}^{\mu} +(1 + \xi){\overline{M}\over 2}n^\mu + {\Delta_\perp^\mu \over 2} \nonumber \\
q_1^{\mu}&=& -2 \xi\tilde{n} ^\mu + {Q^2\over 4 \xi}n^\mu \nonumber \\
k^\mu &=& \left(k\cdot n\,\right)\, \tilde{n}^\mu + \left(k\cdot \tilde{n}\right)\,\,  n^\mu + k_\perp^\mu \nonumber \\ 
\overline{M}^2 &=&M^2 -\frac{\Delta^2}{4}
\end{eqnarray}

with $\bar{P}^2=\overline{M}^2$. We will also use the notation $-q_1^2=Q^2$ for the invariant mass 
of the virtual $Z$ boson and we will denote by $\bar{q}$ the average gauge bosons momenta respectively. 
 
After a collinear expansion of the loop momentum we obtain

\ba
&&T_{A}^{\mu \nu} = i \int \frac{d^4 k}{(2 \pi)^4} \tr\left\{ g_{u}\gamma^\nu \rlap/{P_D}\gamma^{\mu}\tilde{g}\left(U_v-\gamma^5\right)\underline{M}^{u}(k)+\right.\nonumber\\
&&\hspace{3.5cm}\left.g_{d}\gamma^\nu \rlap/{P_D}\gamma^{\mu}\tilde{g}\left(D_v-\gamma^5\right)\underline{M}^{d}(k)\right\} \nonumber\\
&&T_{B}^{\mu \nu} = i \int \frac{d^4 k}{(2 \pi)^4} \tr\left\{ \tilde{g}\gamma^\mu \left(U_v -\gamma^5\right)\rlap/{P_E}\gamma^{\nu} g_u \underline{M}^{u}(k)+\right.\nonumber\\
&&\hspace{3.5cm}\left.\tilde{g}\gamma^\mu \left(D_v -\gamma^5\right)\rlap/{P_E}\gamma^{\nu} g_d \underline{M}^{d}(k)\right\}
\label{coll}
\ea

where we have used the following notations

\ba
&&g_u = \frac{2}{3}e,\hspace{1cm}g_d=\frac{1}{3}e,\hspace{1cm}\tilde{g}=\frac{g}{4 \cos{(\theta_{W})}},\nonumber\\
&&U_v=1-\frac{8}{3} \sin^2\theta_{W},\hspace{1cm}D_v=1-\frac{4}{3}\sin^2 \theta_{W},\nonumber\\\\
&&\rlap/{P_D}=\frac{\ds{k}-\alpha\ds{\Delta}+\ds{q_1}}{\left(k - \alpha\Delta +q_1\right)^2 + i \epsilon}\,\,,\nonumber\\
&&\rlap/{P_E}=\frac{\ds{k}-\ds{q_1}+\ds{\Delta}(1-\alpha)}{\left(k - q_1 +\Delta(1-\alpha)\right)^2 +i\epsilon} \,\,
\ea

where the constant $\alpha$ ($\alpha$ is a free parameter) ranges between $0$ and $1$.
The $\underline{M}$ matrix is the quark density matrix and is given by

\begin{eqnarray}
\underline{M}_{ab}^{(i)}(k) &=& \int d^4y e^{i k\cdot y} \langle P'|
\overline{\psi}_{a}^{(i)}(-\alpha y) \psi_{b}^{(i)}((1-\alpha)y) |P \rangle.
\label{matrix}
\end{eqnarray}

The index $i=u,d$ runs on flavours.
Using a Sudakov decomposition of the internal loop we can rewrite $T_{A}^{\mu\nu}$ and $T_{B}^{\mu\nu}$  as

\ba
&&T_{A}^{\mu\nu}=i\,\int \frac{d( k\cdot n)}{(2 \pi)^4}d(k\cdot \tilde{n}) d^2 k_{\perp}\int\frac{d\lambda}{(2\pi)}dz e^{i\lambda (z-k\cdot n)}\nonumber\\
&&\;\;\;\;\;\;\;\;\;\;\;\;\tr\left\{ g_{u}\gamma^\nu \rlap/{P_D}\gamma^{\mu}\tilde{g}\left(U_v-\gamma^5\right)\underline{M}^{u}(k)+\right.\nonumber\\
&&\hspace{2cm}\left.g_{d}\gamma^\nu \rlap/{P_D}\gamma^{\mu}\tilde{g}\left(D_v-\gamma^5\right)\underline{M}^{d}(k)\right\} \nonumber\\\\
&&T_{B}^{\mu\nu}=i\,\int \frac{d( k\cdot n)}{(2 \pi)^4}d(k\cdot \tilde{n}) d^2 k_{\perp}\int\frac{d\lambda}{(2\pi)}dz e^{i\lambda (z-k\cdot n)}\nonumber\\
&&\;\;\;\;\;\;\;\;\;\;\;\;\tr\left\{ \tilde{g}\gamma^\mu \left(U_v -\gamma^5\right)\rlap/{P_E}\gamma^{\nu} g_u \underline{M}^{u}(k)+\right.\nonumber\\
&&\hspace{2cm}\left.\tilde{g}\gamma^\mu \left(D_v -\gamma^5\right)\rlap/{P_E}\gamma^{\nu} g_d \underline{M}^{d}(k)\right\}
\ea

to which we will refer as the direct and the exchange diagram respectively.
It is also convenient to introduce two new linear combinations $T^{\mu\nu}=T_{A}^{\mu\nu}+T_{B}^{\mu\nu}=\tilde{T}_{A}^{\mu\nu}+\tilde{T}_{B}^{\mu\nu}$ which will turn useful in order to separate Vector (V) and axial vector parts (A) of the expansion

\ba
&&\tilde{T}_{A}^{\mu\nu}=i\int \frac{d( k\cdot n)}{(2 \pi)^4}d(k\cdot\tilde{n}) d^2 k_{\perp}\int\frac{d\lambda}{(2\pi)}dz e^{i\lambda (z-k\cdot n)}\nonumber\\
&&\;\;\;\;\;\;\;\;\;\;\;\;\tilde{g} g_{u}U_v \tr\left\{\left[\gamma^\nu \rlap/{P_D}\gamma^{\mu}+\gamma^\mu \rlap/{P_E}\gamma^{\nu}\right]\underline{M}^{u}(k)\right\}+\nonumber\\
&&\hspace{1.4cm}\tilde{g} g_{d}D_v\tr\left\{\left[\gamma^\nu \rlap/{P_D}\gamma^{\mu}+\gamma^\mu \rlap/{P_E}\gamma^{\nu}\right]\underline{M}^{d}(k)\right\} \nonumber\\\\
&&\tilde{T}_{B}^{\mu\nu}=-i\int \frac{d( k\cdot n)}{(2 \pi)^4}d(k\cdot \tilde{n}) d^2 k_{\perp}\int\frac{d\lambda}{(2\pi)}dz e^{i\lambda (z-k\cdot n)}\nonumber\\
&&\;\;\;\;\;\;\;\;\;\;\;\;\tilde{g} g_{u}\tr\left\{\left[\gamma^\nu \rlap/{P_D}\gamma^{\mu}+\gamma^\mu \rlap/{P_E}\gamma^{\nu}\right]\gamma^5\underline{M}^{u}(k)\right\}+\nonumber\\
&&\hspace{1.4cm}\tilde{g} g_{d}\tr\left\{\left[\gamma^\nu \rlap/{P_D}\gamma^{\mu}+\gamma^\mu \rlap/{P_E}\gamma^{\nu}\right]\gamma^5\underline{M}^{d}(k)\right\}.
\ea

with $\tilde{T}_A$ including the vector parts $(V\times V  \,\,+\,\, A\times A)$ and $\tilde{T}_B$ the axial-vector parts 

$(V\times A \,\,+ \,\,A\times V)$. After some algebraic manipulations we finally obtain

\ba
&&\tilde{T}_{A}^{\mu\nu}=\frac{i}{2}\sum_{i=u,d}\tilde{g}g_i C_i\int\frac{d\lambda dz}{(2\pi)}e^{i\lambda z} \left\{\left(\tilde{n}^{\mu} n^{\nu}+\tilde{n}^{\nu} n^{\mu} -g^{\mu\nu} \right) \alpha(z) \langle P'|\overline {\psi}^{(i)}\left(-\frac{\lambda n}{2}\right)\ds{n}\psi^{(i)}\left(\frac{\lambda n}{2}\right)|P\rangle \right.\nonumber\\
&&\hspace{5 cm}\left.+i\epsilon^{\mu\nu\alpha\beta}\tilde{n}_{\alpha} n_{\beta}\beta(z) \langle P'|\overline{\psi}^{(i)}\left(-\frac{\lambda n}{2}\right)\gamma^5 \ds{n}\psi^{(i)}\left(\frac{\lambda n}{2}\right)|P\rangle \right\}\nonumber\\
&&\tilde{T}_{B}^{\mu\nu}=-\frac{i}{2}\sum_{i=u,d}\tilde{g} g_i\int\frac{d\lambda dz}{(2\pi)}e^{i\lambda z}\left\{\left(\tilde{n}^{\mu} n^{\nu}+\tilde{n}^{\nu} n^{\mu} -g^{\mu\nu} \right) \alpha(z)\langle P'|\overline{\psi}^{(i)}\left(-\frac{\lambda n}{2}\right)\gamma^5 \ds{n}\psi^{(i)}\left(\frac{\lambda n}{2}\right)|P\rangle \right.\nonumber\\
&&\hspace{5cm}\left.+i\epsilon^{\mu\nu\alpha\beta}\tilde{n}_{\alpha} n_{\beta} \beta(z)\langle P'|\overline{\psi}^{(i)}\left(-\frac{\lambda n}{2}\right)\ds{n}\psi^{(i)}\left(\frac{\lambda n}{2}\right)|P\rangle \right\} \nonumber\\
\ea

where $C_i=U_v,\,D_v$ and 

\ba
\alpha(z)=\left(\frac{1}{z-{\xi}+i\epsilon} + \frac{1}{z+{\xi}-i\epsilon}\right),\hspace{1.5cm}
\beta(z)=\left(\frac{1}{z-{\xi} +i\epsilon} - \frac{1}{z+{\xi}-i\epsilon}\right)
\ea
are the ``first order'' propagators appearing in the factorization of the amplitude. We recall, if not obvious, that differently from 
DIS, DVCS undergoes factorization directly at amplitude level \cite{CollinsFreund}. 
 
The parameterizations of the non-forward light cone correlators in terms of GPD's is of the form given by Ji at leading twist \cite{Ji1}

\ba  
&&\int\frac{d\lambda }{(2\pi)}e^{i\lambda z}\langle P'|\overline {\psi}\left(-\frac{\lambda n}{2}\right)\gamma^\mu \psi\left(\frac{\lambda n}{2}\right)|P\rangle=\nonumber\\\nonumber\\
&&H(z,\xi,\Delta^2)\overline{U}(P')\gamma^\mu U(P) + E(z,\xi,\Delta^2)\overline{U}(P')\frac{i\sigma^{\mu\nu} \Delta_{\nu}}{2M} U(P) + .....\nonumber\\
&&\int\frac{d\lambda }{(2\pi)}e^{i\lambda z}\langle P'|\overline {\psi}\left(-\frac{\lambda n}{2}\right)\gamma^\mu \gamma^5 \psi\left(\frac{\lambda n}{2}\right)|P\rangle=\nonumber\\\nonumber\\
&&\tilde{H}(z,\xi,\Delta^2)\overline{U}(P')\gamma^\mu\gamma^5 U(P) + \tilde{E}(z,\xi,\Delta^2)\overline{U}(P')\frac{\gamma^5 \Delta^{\mu}}{2M}U(P) + .....
\ea

which have been expanded in terms of functions $H,E, \tilde{H}, \tilde{E}$ \cite{Ji2} and the ellipses are meant to denote the higher-twist contributions. 
It is interesting to observe that the amplitude is still described by the same light-cone correlators as in the electromagnetic case (vector, axial vector) but now parity is not conserved.

\section{Operatorial analysis}

The operatorial structure of the T-order product of one electroweak current and one electromagnetic 
current is relevant in order 
to identify the independent amplitudes appearing in the correlator at leading twist and the study is presented here. 
We will identify four operatorial structures. For this purpose  let's start from the Fourier transform of the correlator of the two currents
\ba
T_{\mu \nu}=i\int d^{4}x e^{i qx} \langle P_2|T\left(J_{\nu}^{\gamma}(x/2) J_{\mu}^{Z_{0}}(-x/2)\right)|P_1\rangle\,, 
\ea
where for the neutral and electromagnetic currents we have the following expressions
\ba
\label{currents}
&&J^{\mu Z_{0}}(-x/2)=\frac{g}{2 \cos{\theta_W}}\overline{\psi}_{u}(-x/2)\gamma^{\mu}(g^{Z}_{u V}+g^{Z}_{u A}\gamma^5)\psi_{u}(-x/2)+\overline{\psi}_{d}(-x/2)\gamma^{\mu}(g^{Z}_{d V}+g^{Z}_{d A}\gamma^5)\psi_{d}(-x/2),\,\nonumber\\
&&J^{\nu, \gamma}(x/2)=\overline{\psi}_{d}(x/2)\gamma^{\nu}\left(-\frac{1}{3}e\right)\psi_{d}(x/2) +\overline{\psi}_{u}(x/2)\gamma^{\nu}\left(\frac{2}{3}e\right)\psi_{u}(x/2)\,.
\ea
By simple calculations one obtains
\ba
\label{timeorder}
&&\langle P_2|T\left(J_{\nu}^{\gamma}(x/2) J_{\mu}^{Z_{0}}(-x/2)\right)|P_1\rangle=\nonumber\\
&&\langle P_2|\overline{\psi}_{u}(x/2)g_u \gamma_{\nu}S(x)\gamma_{\mu}(g^{Z}_{u V}+g^{Z}_{u A}\gamma^5)\psi_{u}(-x/2)-\nonumber\\
&&\hspace{0.8 cm}\overline{\psi}_{d}(x/2)g_d \gamma_{\nu}S(x)\gamma_{\mu}(g^{Z}_{d V}+g^{Z}_{d A}\gamma^5)\psi_{d}(-x/2)+\nonumber\\
&&\hspace{0.8 cm}\overline{\psi}_{u}(x/2)\gamma_{\mu}(g^{Z}_{u V}+g^{Z}_{u A}\gamma^5)S(-x) g_u \gamma_{\nu}\psi_{u}(x/2)-\nonumber\\
&&\hspace{0.8 cm}\overline{\psi}_{d}(x/2)\gamma_{\mu}(g^{Z}_{d V}+g^{Z}_{d A}\gamma^5)S(-x) g_d \gamma_{\nu}\psi_{d}(x/2)|P_1 \rangle\,.\nonumber\\
\ea
The coefficients used in eqs. (\ref{currents}, \ref{timeorder}) $g_V^{Z}$ and $g_A^{Z}$, are 
\ba
&&g^{Z}_{u V}=\frac{1}{2} + \frac{4}{3} \sin^{2}\theta_{W}\hspace{1.5 cm}g^{Z}_{u A}=-\frac{1}{2}\nonumber\\
&&g^{Z}_{d V}=-\frac{1}{2} + \frac{2}{3} \sin^{2}\theta_{W}\hspace{1.5 cm}g^{Z}_{d A}=\frac{1}{2}\,,
\ea
and 
\ba
g_{u}=\frac{2}{3},&&g_{d}=\frac{1}{3}
\ea
are the absolute values of the charges of the up and down quarks in units of the electron charge.

The function $S(x)$ denotes the free quark propagator
\ba
S(x)\approx \frac{i \rlap/{x}}{2\pi^2(x^2-i\eps)^2}.
\ea

After some standard identities for the $\gamma$'s products 
\ba
&&\gamma_{\mu}\gamma_{\alpha}\gamma_{\nu}=S_{\mu\alpha\nu\beta}\gamma^{\beta}+i\eps_{\mu\alpha\nu\beta}\gamma^{5}\gamma^{\beta},\nonumber\\
&&\gamma_{\mu}\gamma_{\alpha}\gamma_{\nu}\gamma^{5}=S_{\mu\alpha\nu\beta}\gamma^{\beta}\gamma^{5}-i\eps_{\mu\alpha\nu\beta}\gamma^{\beta},\nonumber\\
&&S_{\mu\alpha\nu\beta}=\left(g_{\mu\alpha}g_{\nu\beta}+g_{\nu\alpha}g_{\mu\beta}-g_{\mu\nu}g_{\alpha\beta} \right),
\ea
we rewrite the correlators as 
\ba
&&T_{\mu\nu}=i\int d^{4}x\frac{e^{i q x} x^{\alpha}}{2\pi^2(x^2 -i\eps)^2}
\langle P_2|\left[g_u g_{u V}\left(S_{\mu\alpha\nu\beta}O^{\beta}_{u} -i\eps_{\mu\alpha\nu\beta}O^{5 \beta}_{u}\right)-g_u g_{u A}\left(S_{\mu\alpha\nu\beta}\tilde{O}^{5 \beta}_{u}-i\eps_{\mu\alpha\nu\beta}\tilde{O}^{\beta}_{u}\right)\right.\nonumber\\
&&\hspace{5.5 cm}\left.-g_d g_{d V}\left(S_{\mu\alpha\nu\beta}O^{\beta}_{d}-i\eps_{\mu\alpha\nu\beta}O^{5 \beta}_{d}\right)+g_d g_{d A}\left(S_{\mu\alpha\nu\beta}\tilde{O}^{5 \beta}_{d} -i\eps_{\mu\alpha\nu\beta}\tilde{O}^{\beta}_{d}\right)\right]|P_1\rangle.\,\nonumber\\
\ea
The $x$-dependence of the operators in the former equations was suppressed. 

Whence the relevant operators are denoted by 
\ba
&&\tilde{O}_{a}^{\beta}(x/2,-x/2)=\overline{\psi}_{a}(x/2)\gamma^{\beta}\psi_{a}(-x/2)+\overline{\psi}_{a}(-x/2)\gamma^{\beta}\psi_{a}(x/2),\nonumber\\
&&\tilde{O}_{a}^{5 \beta}(x/2,-x/2)=\overline{\psi}_{a}(x/2)\gamma^{5}\gamma^{\beta}\psi_{a}(-x/2)-\overline{\psi}_{a}(-x/2)\gamma^{5}\gamma^{\beta}\psi_{a}(x/2),\nonumber\\
&&O_{a}^{\beta}(x/2,-x/2)=\overline{\psi}_{a}(x/2)\gamma^{\beta}\psi_{a}(-x/2)-\overline{\psi}_{a}(-x/2)\gamma^{\beta}\psi_{a}(x/2),\nonumber\\
&&O_{a}^{5 \beta}(x/2,-x/2)=\overline{\psi}_{a}(x/2)\gamma^{5}\gamma^{\beta}\psi_{a}(-x/2)+\overline{\psi}_{a}(-x/2)\gamma^{5}\gamma^{\beta}\psi_{a}(x/2)\,,\nonumber\\
\ea
where $a$ is a flavour index.

\section{Phases of the Amplitude}

The numerical computation of the cross section requires a prescription for a correct handling of the singularities in the integration region at $z=\pm\xi$. 
The best way to proceed is to work out explicitly the structure of the factorization formula of the amplitude using the Feynman prescription for going around the singularities, thereby isolating a principal value integral (P.V., which is real) and an imaginary contribution coming from the $\delta$ function term. A  P.V. integral is expressed in terms of ``plus'' distributions and of logarithmic terms, as illustrated below. 
The expression of the factorization formula of the process in the parton model, in which $\alpha(z)$ and $\beta(z)$ appear as factors in the coefficient functions, is then given by
\ba 
&&\hspace{4cm}{\cal{M}}_{f i}=J^{\mu}_{Z}(q_1)D(q_1)\eps^{\nu *}(q_1 -\Delta)\nonumber\\\nonumber\\
&&\times\left\{\frac{i}{2}\tilde{g}g_u U_v \int_{-1}^{1}dz\left(\tilde{n}^{\mu}n^{\nu}+\tilde{n}^{\nu}n^{\mu}-g^{\mu\nu} \right)\right.\nonumber\\
&&\left.\alpha(z)\left[H^u (z,\xi,\Delta^{2})\overline{U}(P_2)\ds{n}U(P_1) + E^u(z,\xi,\Delta^2)\overline{U}(P_2)\frac{i\sigma^{\mu\nu}n_{\mu}\Delta_{\nu}}{2 M}U(P_1)\right]+\right.\nonumber\\
&&\left.\beta(z)i\eps^{\mu\nu\alpha\beta}\tilde{n}_{\alpha}n_{\beta}\left[\tilde{H}^u (z,\xi,\Delta^{2})\overline{U}(P_2)\ds{n}\gamma^{5}U(P_1)+\tilde{E}^{u}(z,\xi,\Delta^{2})\overline{U}(P_2)\gamma^{5}\left(\Delta\cdot n\right)U(P_1)\right]+\right.\nonumber\\
&&\left.\frac{i}{2}\tilde{g}g_d D_v\int_{-1}^{1}dz\left\{u\rightarrow d\right\}-\right.\nonumber\\\nonumber\\
&&\left.\frac{i}{2}\tilde{g}g_u\int_{-1}^{1}dz\left(-\tilde{n}^{\mu}n^{\nu}-\tilde{n}^{\nu}n^{\mu}+ g^{\mu\nu}\right)\right.\nonumber\\
&&\left.\alpha(z)\left[\tilde{H}^u (z,\xi,\Delta^{2})\overline{U}(P_2)\ds{n}\gamma^{5}U(P_1) + \tilde{E}^u(z,\xi,\Delta^2)\overline{U}(P_2)\frac{i \gamma^{5}\Delta\cdot n}{2 M}U(P_1)\right]+\right.\nonumber\\
&&\left.\beta(z)i\eps^{\mu\nu\alpha\beta}\tilde{n}_{\alpha}n_{\beta}\left[H^u (z,\xi,\Delta^{2})\overline{U}(P_2)\ds{n}U(P_1)+E^{u}(z,\xi,\Delta^{2})\overline{U}(P_2)\frac{i\sigma^{\mu\nu}n_{\mu}\Delta_{\nu}}{2 M}(P_1)\right]-\right.\nonumber\\
&&\left.\frac{i}{2}\tilde{g}g_d\int_{-1}^{1}dz\left\{u\rightarrow d\right\}\right\}.\nonumber\\
\ea

To handle the singularity on the path of integration in the factorization formula, as we have already mentioned, 
we use the Feynman ($i\epsilon$) prescription, thereby  generating imaginary parts. 
In particular, any standard integral containing imaginary parts is then separated into real and imaginary contributions as 

\beq
\int dz {T(z)\over z \mp \xi \pm i \epsilon}= PV \int_{-1}^1 dz {T(z)\over z \mp\xi} \,\mp \,i \pi T(\pm \xi)
\eeq

for a real coefficient $T(z)$. 
We then rewrite the P.V. integral in terms of ``plus'' distributions 

\ba
P.V.\int_{-1}^{1} dz {H(z)\over z - \xi}&=&
\int_{-1}^{1}dz \frac{H(z) - H(\xi)}{z - \xi}+ H(\xi)
\log\left({1-\xi\over 1 +\xi}\right) \nonumber \\
&=& \int_{-1}^1 dz {\cal Q}(z)H(z) + 
\int_{-1}^1 dz \bar{{\cal Q}}(z)H(z) + H(\xi)\log\left({1-\xi\over 1 +\xi}\right) \nonumber \\
\label{rex1}
\ea

where 

\ba
{\cal Q}(z)&=&\theta(-1\leq z\leq \xi) {1\over (z - \xi)_+}\nonumber \\
&=&\theta(-1\leq z\leq \xi)\left( {\theta(z < \xi)\over (z - \xi)} -
\delta(z- \xi)\int_{-1}^\xi{dz\over (z - \xi)}\right)\nonumber \\
\bar{\cal Q}(z)&=&\theta(\xi\leq z\leq 1) {1\over (z - \xi)_+}\nonumber \\
&=&\theta(-1\leq z\leq \xi)\left( {\theta(z> \xi)\over (z - \xi)} -
\delta(z- \xi)\int_{\xi}^1{dz\over (z - \xi)}\right)\nonumber \\
\label{rex2}
\ea

and the integrals are discretized using finite elements methods, in order to have high numerical accuracy. 
This last point is illustrated in Appendix C, where the computations are done analytically on a grid and then the grid spacing is sent to zero. 

We can now proceed and compute the cross section. We define the scalar amplitude

\ba  
{\cal M}_{f i}=J_{Lep}^{\mu}(q_1)D(q_1)T^{\mu\nu}\epsilon^{*\nu}(q_1 -\Delta)
\ea

where $D(q_1)$ is the $Z_{0}$ propagator in the Feynman gauge and $J_{Lep}^{\mu}(q_1)$ is the leptonic current 
and we have introduced the polarization vector for the final state photon $\epsilon^\nu$.

In particular, for the squared amplitude we have

\ba
|{\cal M}_{f i}|^2 = -L^{\mu\lambda}D(q_1)^2 T_{\mu\nu}T^{*\nu}_{\lambda}\,
\ea
which is given, more specifically, by

\beq
{\cal |M|}^2 =\int_{-1}^1 dz \int_{-1}^1 dz' \left( K_1(z,z') \alpha(z) \alpha^*(z') + K_2(z,z') \beta(z) \beta^*(z')  \right)       
\label{Fact1}
\eeq

with $K_1$ and $K_2$ real functions, combinations of the generalized distributions $(H,\tilde{H}, E,\tilde{E})$ with appropriate kinematical factors. Mixed contributions proportional to $\alpha(z)\beta^*(z')$ and $\beta(z)\alpha^*(z')$ cancel both in their real and imaginary parts and as such do not contribute to the phases. A similar result holds also for the pure electromagnetic case.

After some further manipulations, we finally rewrite the squared amplitude in terms of a P.V. contribution plus some additional 
terms coming from the imaginary parts 

\ba
{\cal |M|}^2 &=& P.V.\int_{-1}^1 dz \int_{-1}^1 dz' \left( K_1(z,z') \alpha(z) \alpha^*(z') + K_2(z,z') \beta(z) \beta^*(z')  \right)\nonumber \\       
&&+ \pi^2 \left( K_1(\xi,\xi) - K_1(\xi,-\xi) - K_1((-\xi,\xi) + K_1(-\xi,-\xi)\right)\nonumber \\
&&+ \pi^2 \left( K_2(\xi,\xi) + K_2(\xi,-\xi) + K_2((-\xi,\xi) + K_2(-\xi,-\xi)\right)
\ea
which will be analized numerically in the sections below. In order to proceed with the numerical result, it is necessary 
to review the standard construction of the nonforward parton distribution functions in terms of the forward distributions, 
which is the topic of the next section.  

\section{Construction of the Input Distributions}

The computation of the cross section proceeds rather straightforwardly, though the construction of the initial conditions is more involved compared to the forward (DIS) case. 
This construction has been worked out in several papers \cite{bel1,radmus,Biernat,Freu&McD2,Initialcond,Rad1,radmus} 
in the case of standard DVCS, using a diagonal input appropriately extended to the non-diagonal kinematics. Different types of nonforward parton distribution, all widely used in the numerical implementations have been put forward, beside Ji's original distributions, which we will be using in order to construct the initial conditions for our process.

For our purposes it will be useful to introduce Golec-Biernat and Martin's (GBM) distributions \cite{Biernat} 
at an intermediate step, which are linearly related to Ji's distributions.   

\begin{figure}[t]
{\par\centering \resizebox*{10cm}{!}{\includegraphics{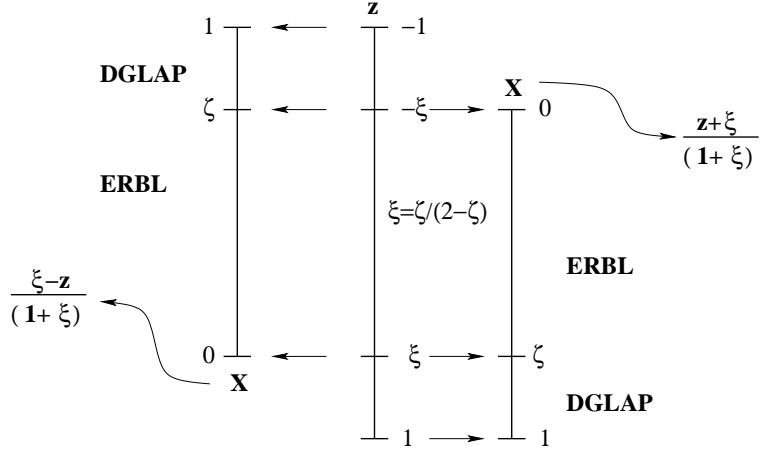}} \par}
\caption{The relationship between ${\cal F}^q (X,\zeta)$, ${\cal F}^{\bar q} (X,\zeta) $ and Ji's function $H^q (z,\xi)$.}
\label{relatio}
\end{figure}

We recall, at this point, that the quark distributions $H_{q}(z,\xi)$ have support in $z\in [-1,1]$, describing 
both quark and antiquark distributions for $z>0$ and $z<0$ respectively. In terms of GBM distributions, 
two distinct distributions ${\hat{\cal{F}}}^{\bar{q}}(X,\zeta)$ and ${\hat{\cal{F}}}^{q}(X,\zeta)$ 
with $0\leq X\leq 1$ are needed in order to cover the same information contained in Ji's distributions using only a positive 
scaling variable ($X$).  
 In the region $X\in (\zeta,1]$ the functions ${\hat{\cal{F}}}^{q}$ and ${\hat{\cal{F}}}^{\bar{q}}$ are independent, but if $X\le\zeta$ they are related to each other, as shown in the (by now standard) plot in Fig.~\ref{relatio}.

In this new variable ($X$) the DGLAP region is described by $X > \zeta$ ($|z| > \xi$), and the ERBL region by $X<\zeta$ ($|z| < \xi$). 
In the ERBL region, $\hat{{\cal F}}^q$ and $\hat{{\cal F}}^{\bar q}$ are not independent.

The relation between $H(z,\xi)$ and $\hat{{\cal{F}}}^{q}(X,\zeta)$ can be obtained explicitly \cite{Freund1} as follows: 
for $z \in [-\xi,1]$ we have

\begin{equation}
\hat{{\cal F}}^{q,i} \left(X = \frac{z+\xi}{1 + \xi},\zeta\right) = \frac{H^{q,i} (z,\xi)}{1-\zeta/2} \, ,
\label{curlyq}
\end{equation}

and for $z \in [-1,\xi]$

\begin{equation}
\hat{{\cal F}}^{{\bar q},i} \left(X = \frac{\xi -z}{1 + \xi},\zeta\right) = -\frac{H^{q,i} (z,\xi)}{1-\zeta/2} \, .
\label{curlyqbar}
\end{equation}

where $i$ is a flavour index.

In our calculations we use a simplified model for the GPD's where the $\Delta^2$ dependence can be factorized as follows
\cite{Vander&guich,bel1}
\ba
H^{i}(z,\xi,\Delta^2,Q^2)&=&F^{i}_{1}\,(\Delta^2)q^{i}(z,\xi,Q^2)\nonumber\\
\tilde{H}^{i}(z,\xi,\Delta^2,Q^2)&=&G^{i}_{1}(\Delta^2)\,\Delta{q}^{i}(z,\xi,Q^2)\nonumber\\
E^{i}(z,\xi,\Delta^2,Q^2)&=&F^{i}_{2}(\Delta^2)\,r^{i}(z,\xi,Q^2)
\label{deltadropping}
\ea

where $q^{i}(z)$ and $\Delta q^{i}(z)$ are obtained from the standard non-polarized and longitudinally polarized (forward) quark distributions using a specific diagonal ansatz \cite{gluck&reya&vogt98}. 

The ansatz $r^{i}(z,\xi)=q^{i}(z,\xi)$ is also necessary in order for the quark sum rule to hold \cite{Ji3}. 
Analogously, in the case of the $\tilde{E}^{i}$ distributions \cite{Rad1,Man1,Fran1} one can use the 
special model 

\ba
\tilde{E}^{u}=\tilde{E}^{d}=\frac{1}{2\xi} \theta(\xi - |z|)\phi_{\pi}(z/\xi)g_{\pi}(\Delta^2), && g_{\pi}(\Delta^2)=\frac{4 g_{A}^{(3)}M^2}{m_{\pi}^2 -\Delta^2},\;\;\;\;\;\;\;\phi_{\pi}(x)=\frac{4}{3}(1-x^2)\nonumber\\
\ea
valid at small $\Delta^2$,
where $g_{A}^{(3)}=1.267$, $M$ is the nucleon mass and $m_{\pi}$ is the pion mass, with the normalization

\ba
F^{i}_{1}(0)=G^{i}_{1}(0)=1.
\ea

Notice that, analogously to the $H$ distributions, $q^{i}(z,\xi,Q^2)$ and $\Delta q^{i}(z,\xi,Q^2)$, which describe the 
$\Delta^2=0$ limit of the H-distributions, have support in $[-1,1]$ and, again, 
they describe quark distributions (for $z>0$) and antiquark distributions (for $z<0$)

\ba
&&\bar{q}^{i}(z,\xi,Q^2)=- q^{i}(-z,\xi,Q^2)\nonumber\\
&&\Delta \bar{q}^{i}(z,\xi,Q^2)=\Delta q^{i}(-z,\xi,Q^2).
\ea

Now we're going to estabilish a connection between the $q(z,\xi,\bar{Q}^2)$ and the $\hat{{\cal{F}}}^q(X,\zeta)$ functions, 
which is done using Radyushkin's nonforward  ``double distributions'' \cite{Rad1}.    
The construction of the input distributions, 
in correspondence of an input scale $Q_0$, is performed following a standard strategy. This consists in generating nonforward double distributions $f(x,y)$ from the forward ones ($f(x)$) using a ``profile function'' $\pi(x,y)$ \cite{Rad3}

\ba
f(y,x)=\pi(y,x)f(x),
\label{doub}
\ea
where we just recall that the $\pi(y,x)$ function can be represented by  
\ba
\pi(y,x)=\frac{3}{4}\frac{[1-|x|]^2 -y^2}{[1-|x|]^3}\,,
\ea
taken to be of an asymptotic shape (see ref.\cite{Rad3,radmus}) for quarks and gluons.
A more general profile is given by 
\begin{equation}
\pi(x,y) = \frac{\Gamma(2b + 2)}{2^{2b+1} \Gamma^2 (b+1)} \frac{[(1 -|x|)^2 - y^2]^b}{(1 -|x|)^{2b+1}}\,\,
\label{profile}
\end{equation}

and normalized so that 

\ba
&&\int^{1-|x|}_{-1+|x|} dy \, \pi(x,y)= 1 \, .
\label{profile2}
\ea

$b$ parameterizes the size of the skewing effects starting from the diagonal input. Other choices of 
the profile function are also possible.
For instance, the double distributions (DD) defined above have to satisfy a symmetry constraint 
based on hermiticity. This demands that these must be symmetric with respect to the exchange $y\longleftrightarrow1-x-y$, and a profile function which respects this symmetry constraint is given by \cite{Piller}
\ba
\label{constraint}
\pi(x,y)=\frac{6y(1-x-y)}{(1-x)^3}\,.
\ea

This symmetry is crucial for establishing proper analytical properties of meson production amplitudes. We will be using below this profile and 
compare the cross section obtained with it against the one obtained with (\ref{profile}).

Now we are able to generate distributions $q(z,\xi,Q^2)$ in the $z$ variable at $\Delta^2=0$, $q(z,\xi,Q^2)$, by integrating over the 
longitudinal fraction of momentum exchange $y$ characteristic of the double distributions 

\ba
q(z,\xi,Q^2)=\int_{-1}^{1}dx'\int_{-1+|x'|}^{1-|x'|}dy'\delta(x'+\xi y'-z)f(y',x',Q^2).
\label{reduction}
\ea

Using (\ref{reduction}) and the expression of the profile functions introduced above, the GBM distributions are generated by the relation
\ba
&&\hat{{\cal F}}^{q,a} (X,\zeta) = \frac{2}{\zeta} \int^{X}_{\frac{X-\zeta}{1-\zeta}} dx'
\pi^q \left (x', \frac{2}{\zeta} (X - x') + x' -1 \right) q^a (x') \, .
\ea

with a similar expression for the anti-quark distributions in the DGLAP region $X > \zeta ~(z < - \xi)$ 

\ba
&&\hat{{\cal F}}^{\bar q,a} (X,\zeta) = \frac{2}{\zeta} \int^{\frac{-X+\zeta}{1-\zeta}}_{-X} dx' \pi^q \left(x', -\frac{2}{\zeta} (X + x') + x' +1 \right) {\bar q}^a(|x'|).
\label{DGLAPqbar}
\ea

\begin{figure}[t]
{\par\centering \resizebox*{10cm}{!}{\rotatebox{-90}{\includegraphics{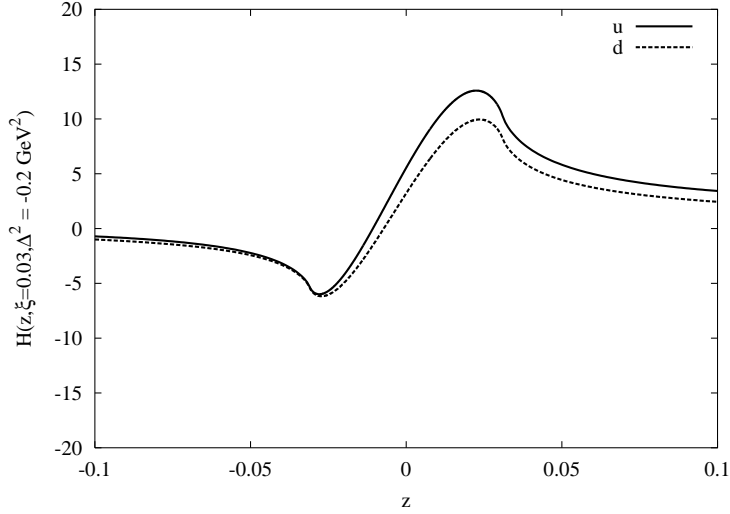}}} \par}
\caption{GPD's $H_u$ and $H_d$ generated by the diagonal parton distribution with a profile function (\ref{profile}) at an initial $0.26$ GeV$^2$}
\label{Hud}
\end{figure}

In the ERBL region, $ X < \zeta ~(|z| < \xi$), after the integration over $y$, we are left with the sum of two integrals

\ba
\hspace{0cm}\hat{{\cal F}}^{q,a} (X,\zeta) = \frac{2}{\zeta} \left[\int^{X}_{0} dx' \pi^q \left(x', \frac{2}{\zeta} (X - x') + x' -1 \right) q^a (x') \, - \right. \nonumber \\
\hspace{0cm}\left. \int^{0}_{X-\zeta} dx' \pi^q \left(x', \frac{2}{\zeta} (X - x') + x' -1 \right) {\bar q}^a (|x'|) \right] \, , \nonumber \\ \\
\hspace{0cm}\hat{{\cal F}}^{{\bar q},a} (X,\zeta) = - \frac{2}{\zeta} \left[ \int^{\zeta - X}_{0} dx' \pi^q \left(x', -\frac{2}{\zeta} (X + x') + x' + 1 \right) q^a (x') \, - \right. \nonumber \\
\hspace{0cm}\left. \int^{0}_{-X} dx' \pi^q \left(x', -\frac{2}{\zeta} (X + x') + x' + 1  \right) {\bar q}^a (|x'|) \right] \, .
\label{erblqqbar1}
\ea

\begin{figure}[t]
{\par\centering \resizebox*{10cm}{!}{\rotatebox{-90}{\includegraphics{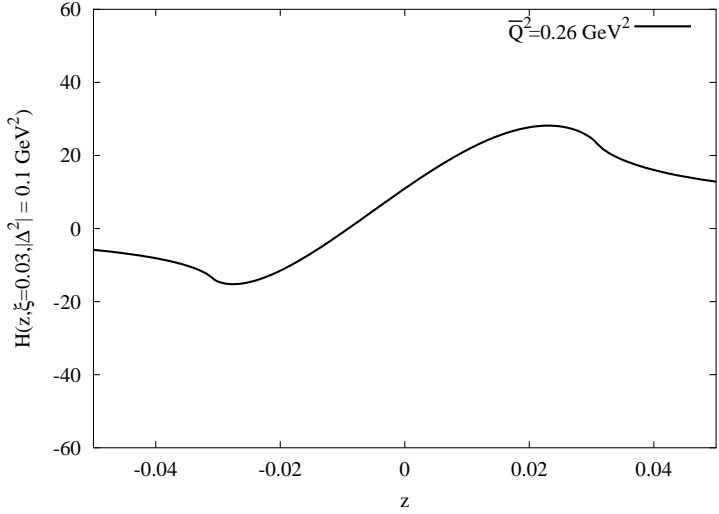}}}\par}
\caption{GPD's flavour singlet combination at $0.26$ GeV$^2$ generated with a profile (\ref{profile})}
\label{HS}
\end{figure}

Solving numerically the integrals we obtain the value of the function ${\cal{F}}^{q}$s on a grid, and using eqs.~(\ref{curlyq}) and \ref{curlyqbar} we end up with the numerical form of the H-distributions. 
We have used diagonal parton distribution functions at $0.26$ GeV$^2$ \cite{gluck&reya&vogt98}
and the results of our numerical implementation can be visualized in Figs.~\ref{Hud} and \ref{HS}.

\section{The Differential Cross Section}
\begin{figure}
{\centering \resizebox*{9cm}{!}{\rotatebox{0}{\includegraphics{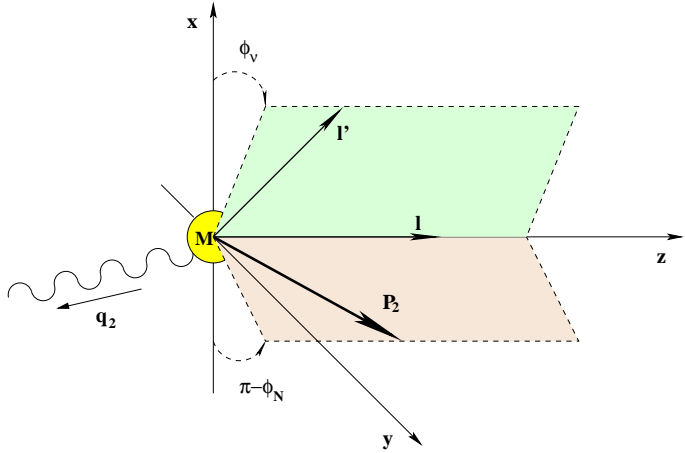}}}\par}
\caption{Kinematics of the process $\nu(l)N(P_1)\rightarrow\nu(l')N(P_2)\gamma(q_2)$}
\label{kinematic.eps}
\end{figure}

Our kinematical setup is illustrated in Fig.~\ref{kinematic.eps}, and we choose momenta in the target frame with the 
following parameterizations

\ba     
l=\left(E,0,0,E\right)\,,&& l'=\left(E',E'\cos{\phi_{\nu}}\sin{\theta_{\nu}},E'\sin{\phi_{\nu}}\sin{\theta_{\nu}},E'\cos{\theta_{\nu}}\right)\,,\nonumber\\
P_1 =\left(M,0,0,0\right)\,,&& P_2 =\left(E_2,|P_2|\cos{\phi_{N}}\sin{\theta_{N}},|P_2|\sin{\phi_{N}}\sin{\theta_{N}},|P_2|\cos{\theta_{N}}\right)\,
\ea
where the incoming neutrino is taken in the positive $\hat{z}$-direction and the nucleon is originally at rest. The first plane 
is identified by the momenta of the final state nucleon and of the incoming neutrino, while the second plane is spanned by the 
final state neutrino and the same $\hat{z}$-axis. $\phi_\nu$ is the angle between the $\hat{x}$ direction and the second plane, while 
$\phi_{N}$ is taken between the plane of the scattered nucleon and the same $\hat{x}$ axis.
We recall that the general form of a differential cross section is given by 

\ba
d\sigma =\frac{1}{4(l\cdot P_1)}|{\cal M}_{fi}|^2 (2\pi)^4 \delta^{(4)} (l+ P_1 -P_2 - l'-q_2 )\frac{d^3 \vec{l'}}{2l^{'0}(2\pi)^3}\frac{d^3 \vec{P_2}}{2P_{2}^{0}(2\pi)^3}\frac{d^3 \vec{q_2}}{2q_{2}^{0}(2\pi)^3}\,\nonumber\\
\ea
and it will be useful to express it in terms of standard quantities appearing in a standard 
DIS process such as Bjorken variable $x$, 
inelasticity parameter $y$, the momentum transfer plus some additional kinematical variables 
typical of DVCS such as the asymmetry parameter $\xi$ and $\Delta^2$. 

\begin{figure}[t]
{\par\centering \resizebox*{10cm}{!}{\rotatebox{0}{\includegraphics{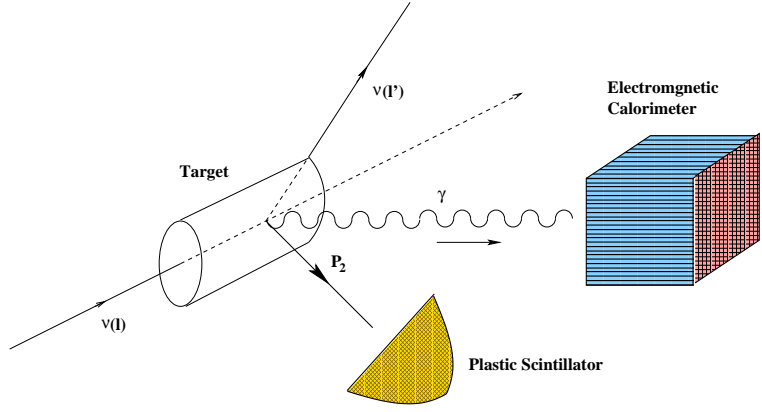}}}\par}
\caption{A pictorial description of the DVNS experimental setup, where the recoiled nucleon is 
detected in coincidence with a final state photon.}
\label{Target.eps}
\end{figure}

We will be using the relations 

\ba
&&\bar{q}^2 = -\frac{1}{2} q_1^2 \left(1-\frac{\Delta^2}{2q_1^2} \right) \approx \frac{1}{2} Q^2\nonumber\\
&&\xi=\frac{x\left(1 -\frac{\Delta^2}{2q_1^2}\right)}{2-x\left(1 -\frac{\Delta^2}{2q_1^2}\right)}\approx \frac{2 x}{2-x}\nonumber\\
\ea
in the final computation of the cross section.
It is also important to note that $\Delta^2$ has to satisfy a kinematical constraint 

\ba
\Delta^{2}_{min}=-\frac{M^2 x^2}{1-x +\frac{x M^2}{Q^2}}\left(1+ {\cal O}\left(\frac{M^2}{Q^2}\right)\right)\,.
\ea
One possible cross section to study, in analogy to the DVCS case \cite{Freu&McD}, is the following    

\ba
\frac{d\sigma}{dx d Q^2 d|\Delta^2| d\phi_{r}}=\frac{y}{Q^2}\frac{d\sigma}{dx dy d|\Delta^2| d\phi_{r}}=\frac{x y^2}{8\pi Q^4}\left(1+ \frac{4M^2 x^2}{Q^2}\right)^{-\frac{1}{2}}|{\cal M}_{fi}|^2.
\ea

where $\phi_{r}$ is the angle between the lepton and the hadron scattering planes. 
To proceed we also need the relations

\ba
&&l\cdot n=\left[\frac{Q^2}{2 x y} -\frac{(1+\xi)}{2\xi}\frac{Q^2}{2}\right]\chi,\nonumber\\
&&l\cdot \tilde{n}=\frac{Q^2}{2\xi}+\frac{Q^2}{4\xi^2}\left[\frac{Q^2}{2xy}-\frac{(1+\xi)}{2\xi}\frac{Q^2}{2}\right]\chi,\nonumber\\
&&n\cdot q_1 = \frac{2x}{x-2},\hspace{0.5cm}\tilde{n}\cdot q_1= Q^2 \frac{(2-x)}{8x},\nonumber\\
\ea

where $\chi$ is given by 

\ba
\chi=\frac{\xi}{\frac{1+\xi}{2} \frac{Q^2}{4\xi} +\frac{\xi(1-\xi)}{2} \frac{\overline{M}^2}{2}}\,\,.
\ea

After some manipulations we obtain a simplified expression for $|{\cal M}_{fi}|^2$, similarly to eq.~(\ref{Fact1})

\ba
|{\cal M}_{fi}|^2 &=& \int_{-1}^1 dz \int_{-1}^1 dz' \left[A_1(z,z',x,t,Q^2)\alpha(z)\alpha^*(z') + A_2(z,z',x,t,Q^2)\beta(z)\beta^*(z')\right]\nonumber\\
& \times &\left[-2 Q^2 \,(4 M^2 - t)\,(x-2)^2 \,(x-1) \,x^2 y + (t-4 M^2)^2 \,(x-1)^2 \,x^4 y^2\right.\nonumber\\ 
&+& \left.Q^4 (x-2)^4 \,(2 - 2 y + y^2)\right]\nonumber\\
&\times&\left[2 M^2 ({M_Z}^2 + Q^2)^2 \, (x-2)^2\, [Q^2 (x-2)^2 - (4 M^2 -t) (-1 + x) x^2]^2 y^2\right]^{-1}       
\label{amplicompact}
\ea

where $A_1(z,z',x,t,Q^2)$ and $A_2(z,z',x,t,Q^2)$ are functions of the invariants of the 
process and of the entire set of GPD's. Their explicit form is given in Appendix D.
As we have already mentioned, the $(z,z')$ integration is be done by using the Feynman prescription to extract the phases and then 
using the distributional identities (\ref{rex1})and (\ref{rex2}). For numerical accuracy we have discretized the final integrals by finite element 
methods, as shown in Appendix C. 

\begin{figure}
{\centering \resizebox*{9cm}{!}{\rotatebox{-90}{\includegraphics{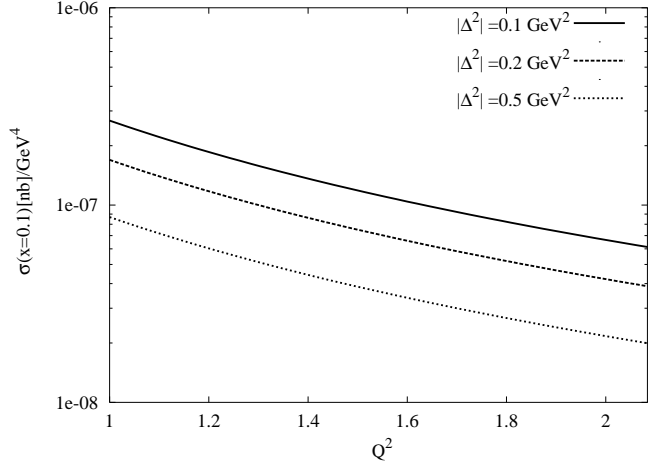}}} \par}
\caption{ DVCS cross section at $x=0.1$ and center of mass energy $M E=10$ GeV$^2$ using the profile (\ref{profile}).}
\label{set2.ps}
\end{figure}

\begin{figure}
{\centering \resizebox*{9cm}{!}{\rotatebox{-90}{\includegraphics{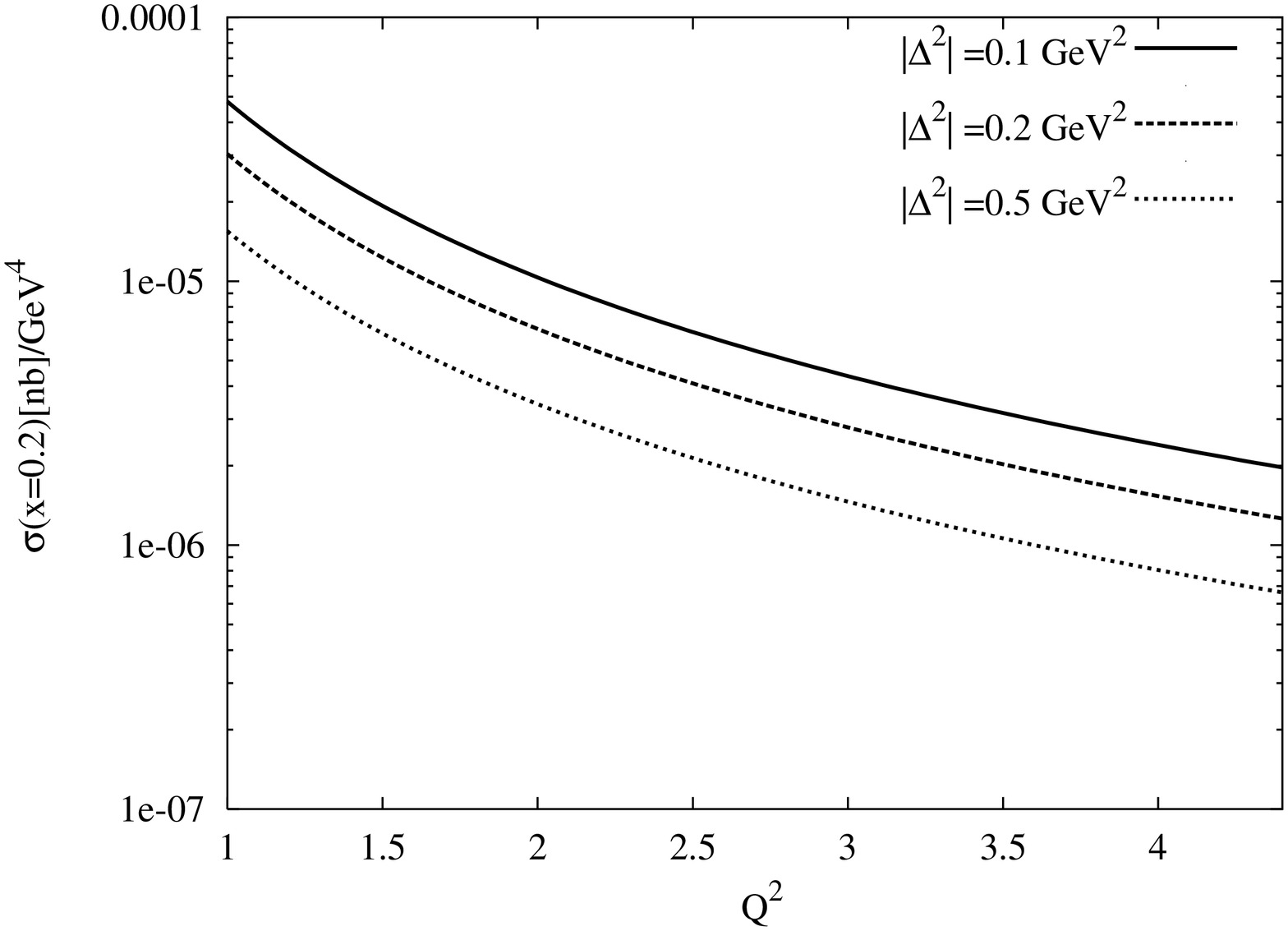}}} \par}
\caption{ DVCS cross section at $x=0.2 $ and center of mass energy $M E=10$ GeV$^2$ using the profile (\ref{profile}).}
\label{set3.ps}
\end{figure}

\begin{figure}
{\centering \resizebox*{9cm}{!}{\rotatebox{-90}{\includegraphics{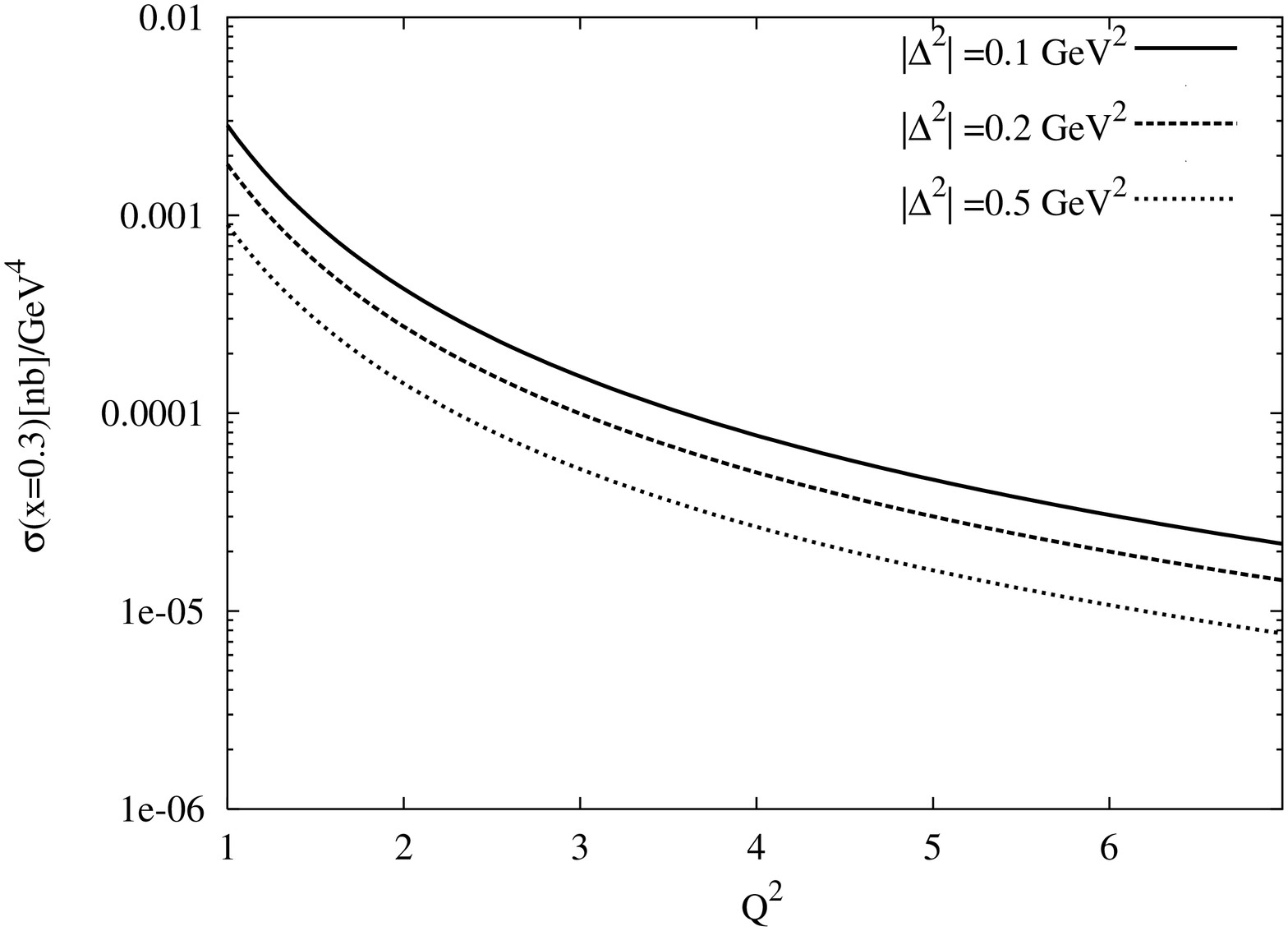}}} \par}
\caption{ DVCS cross section at $x=0.3$ and center of mass energy $M E=10$ GeV$^2$ using the profile (\ref{profile}).}
\label{set4.ps}
\end{figure}

\begin{figure}
{\centering \resizebox*{9cm}{!}{\rotatebox{-90}{\includegraphics{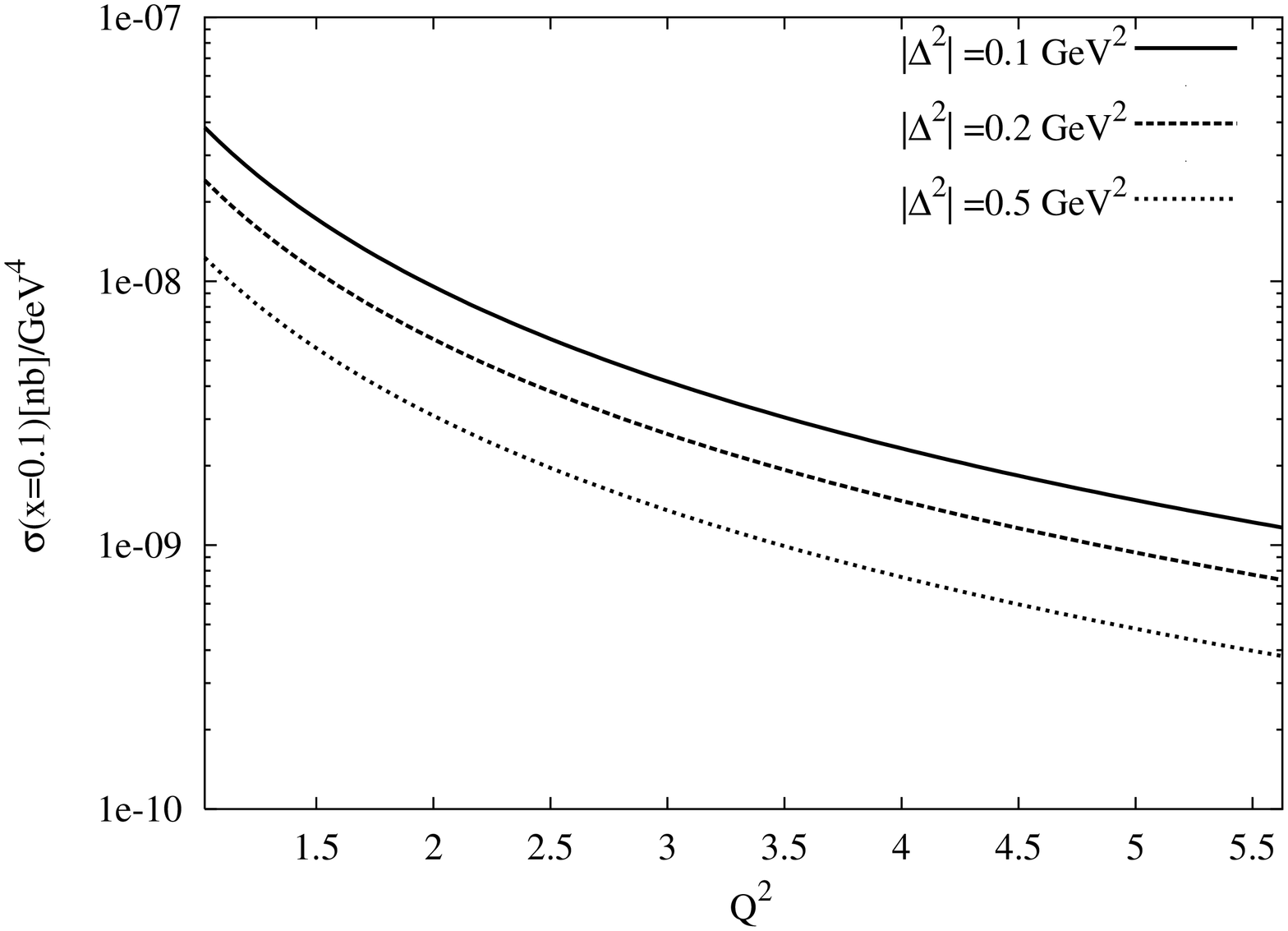}}} \par}
\caption{ DVCS cross section at $x=0.1$ and center of mass energy $M E=27$ GeV$^2$ using the profile (\ref{profile}).}
\label{set5.ps}
\end{figure}

\begin{figure}
{\centering \resizebox*{9cm}{!}{\rotatebox{-90}{\includegraphics{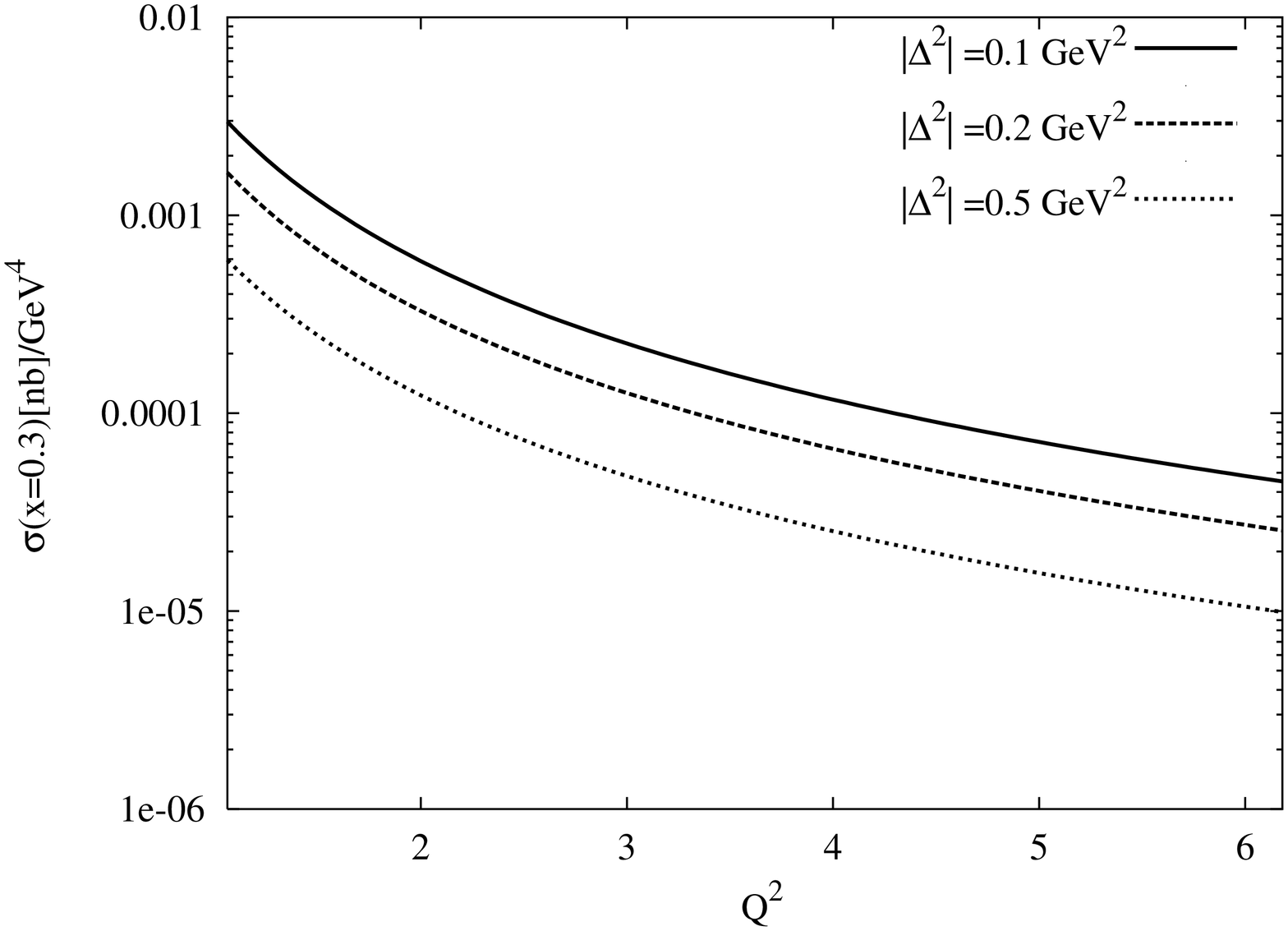}}} \par}
\caption{ DVCS cross section at $x=0.3$ and center of mass energy $M E=10$ GeV$^2$ with NPD functions generated by the profile function 
(\ref{constraint}).}
\label{new_set10.ps}
\end{figure}

\begin{figure}
{\centering \resizebox*{9cm}{!}{\rotatebox{-90}{\includegraphics{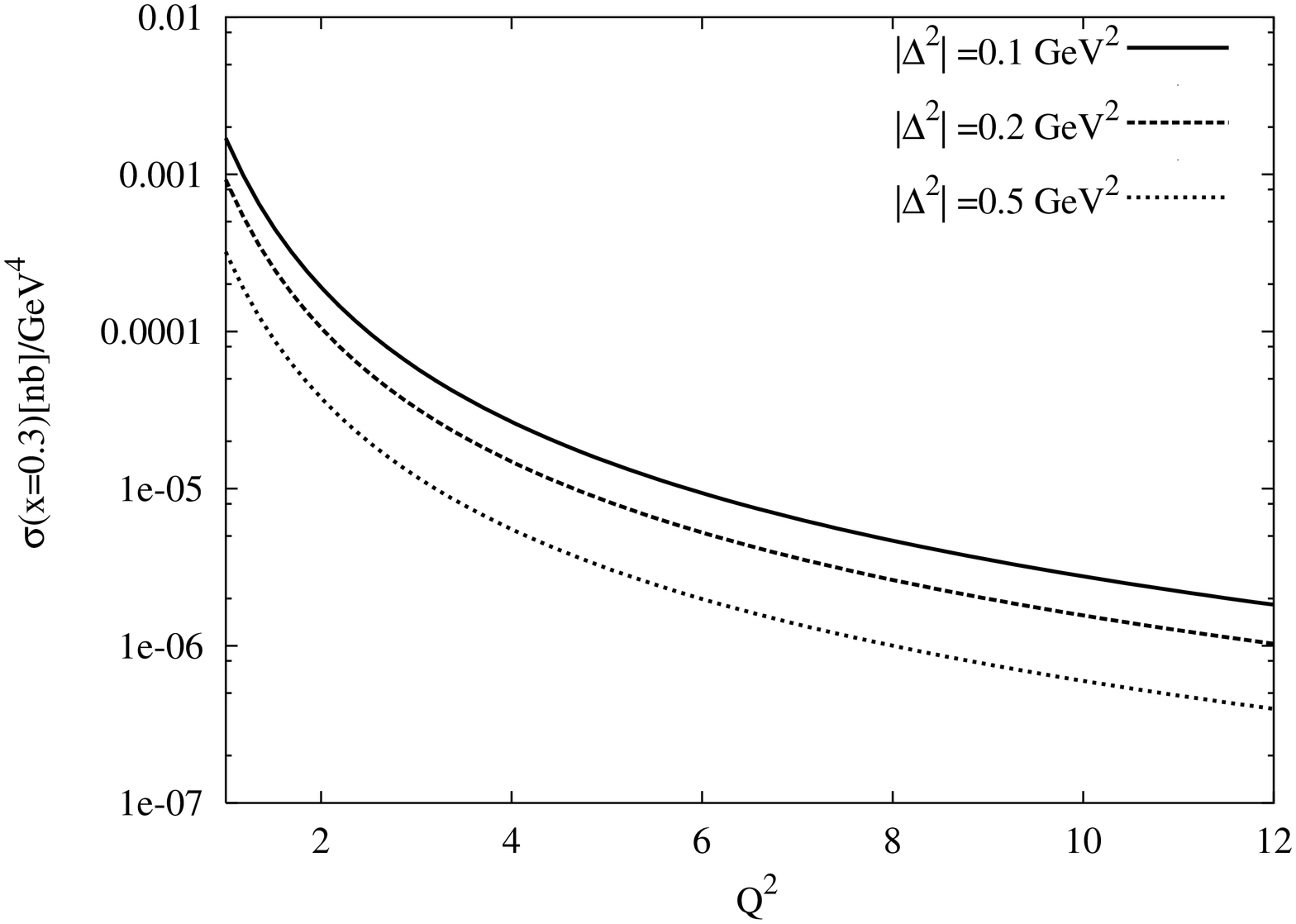}}} \par}
\caption{ DVCS cross section at $x=0.3$ and center of mass energy $M E=27$ GeV$^2$ with NPD functions generated by the profile function 
(\ref{constraint}).}
\label{new_set27.ps}
\end{figure}

\begin{figure}
{\centering \resizebox*{9cm}{!}{\rotatebox{-90}{\includegraphics{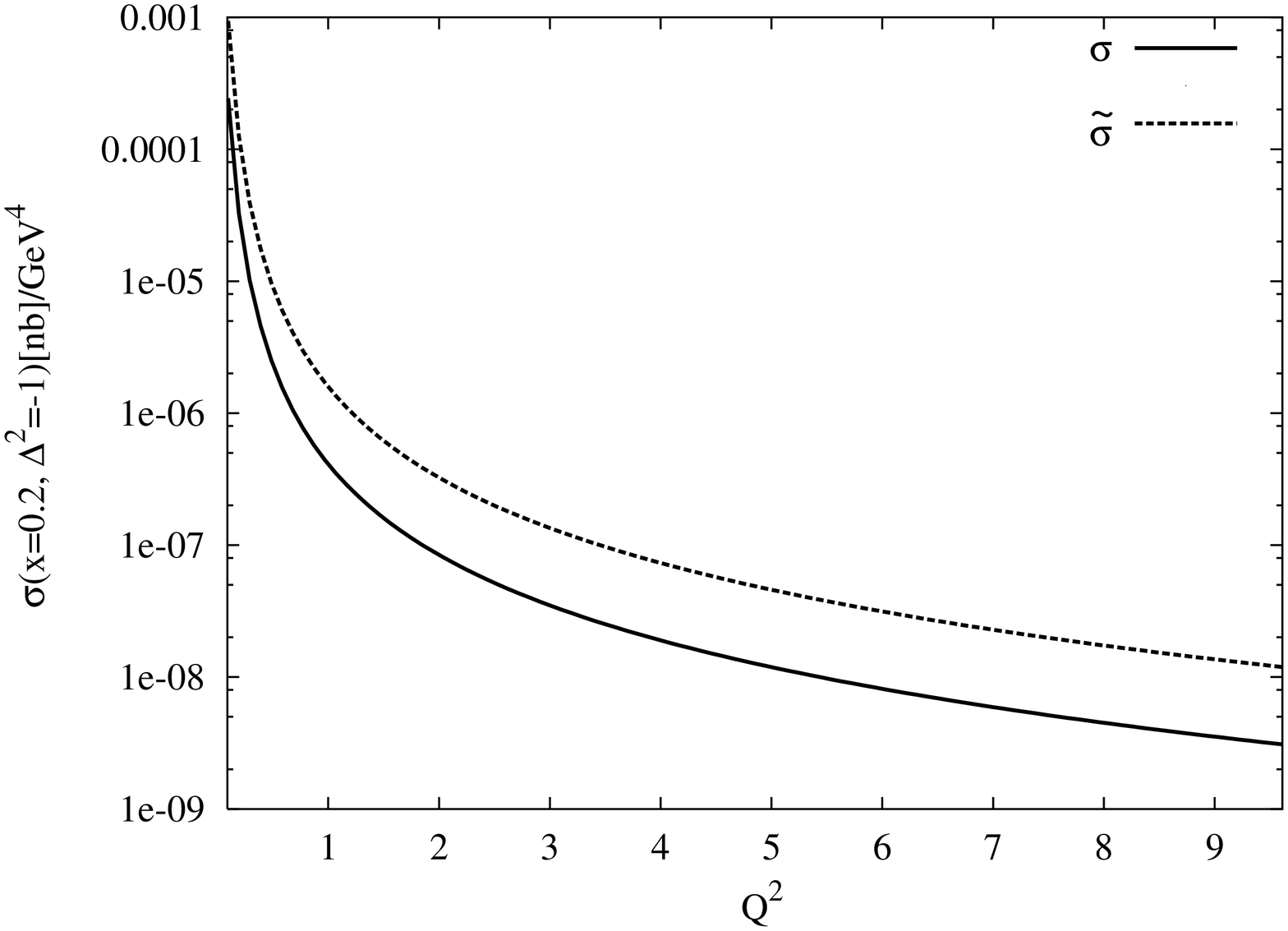}}} \par}
\caption{ DVCS cross sections at $x=0.2$ and center of mass energy $M E=27$ GeV$^2$.
$\tilde{\sigma}$ using profile (\ref{constraint})}.
\label{new_sez.ps}
\end{figure}

We have plotted the differential cross section as a function of $Q^2$, for various values of 
$\Delta^2$ and at fixed $x$ values. We have used both the profile given by (\ref{profile}) (Figs.~\ref{set2.ps}-\ref{set5.ps}) and the 
one given in eq.~(\ref{constraint}) (Figs.~\ref{new_set10.ps} and \ref{new_set27.ps}) and their direct comparison  in a specific kinematical 
region (Fig.~\ref{new_sez.ps}). The different profiles generate differences in the cross sections especially for larger 
$\Delta^2$ values. Notice also that the DVNS cross section decreases rather sharply with $\Delta^2$, at the same time it increases 
appreciably with $x$. The results shown are comparable with other cross sections evaluated in the quasi-elastic region 
($\approx 10^{-5}$ nb) for charged and neutral current interactions, and appear to be sizeable. 
Coherence effects due to neutral current interactions with heavy nuclei, in particular with the neutron component 
may substantially increase 
the size of the cross sections, with an enhancement proportional to $N^2$, where $N$ is the number of neutrons 
\cite{PK}, though an accurate quantification of these effects requires a special study \cite{CCV} which is underway. 
It is worth to emphasize that in the past this contribution 
had never been included in the study of neutrino-nucleon interactions since very little was  known about the intermediate 
energy kinematics in QCD from the point of view of factorization. It seems obvious to us that with the new developments now taking place in the study of QCD 
at intermediate energy, especially in the case of the generalized Bjorken region, of which the deeply virtual scattering limit 
is just a special case, it will be of wide interest to quantify with accuracy the role of these new contributions for neutrino factories. 
In general, one expects that electromagnetic effects are suppressed compared to the standard (hadronic) deeply inelastic cross section, 
and this has led in the past to a parameterization of the intermediate energy cross section as either dominated by the quasi elastic region and/or by the DIS region at higher energies, as we have mentioned in our introduction. 
However, the exclusive cross section has some special positive features, one of them being to provide 
a clean signal  for the detection of weakly interacting particles, 
and we expect that this aspect is going to be of relevance at experimental level.

\section{Conclusions}

We have presented an extension of the standard DVCS process to the case of one neutral current exchange, describing the scattering of a neutrino off a proton in the parton model.
We have described the leading twist behaviour of the cross section; we have found that this is comparable to other typical 
neutrino cross sections and discussed its forward or DIS limit. We have presented a complete formalism for the study 
of these processes in the parton model. The process is the natural generalization of DIS with neutral currents and relies on 
the notion of Generalized Parton Distributions, new constructs in the parton model which have received considerable attention in recent years.
The possible applications of these new processes are manifold and we hope to return in the near future with a discussion of some of the issues not addressed in this work.

\centerline{\bf Acknowledgments}
We warmly thank J.D. Vergados for illuminating discussions on the detectability of this process, and 
Giovanni Chirilli of Old Dominion University for discussions and for collaborating at the derivation of the results of section 5. 
The work of C.C. is supported in part under the grant HPRN-CT-2000-00121. 
C.C. and M.G. thank T.N. Tomaras and 
E. Kiritsis for hospitality at the Univ. of Crete.
C.C. thanks the Theory Group at the University of Zurich at Irchel for hospitality during the last stages of this work.

\section*{Appendix A}

The collinear expansion of the internal loop momentum $k$ allows to identify the light cone operators appearing in the process at leading twist.
We recall at this point that the analysis of the hand-bag contribution is carried out exactly as in the electromagnetic case. 

To perform the collinear expansion and isolate the light-cone correlators of DVNS from the hand-bag contribution 
we use the relation
\begin{eqnarray}
\int \frac{d\lambda \ dx}{2 \pi} e^{i \lambda (x - k\cdot n)} = 1
\end{eqnarray}

inside the expression of $T^{\mu \nu}$ in order to obtain

\begin{eqnarray}
T^{\mu \nu} &=& - \int \frac{d^4k}{(2 \pi)^4} \int \frac{d\lambda \ dx}{2 \pi}
e^{i \lambda (x - k\cdot n)}  \nonumber \\
&& \tr\left\{ \left[ \gamma^\nu
\frac{i}{\rlap/k - \alpha \rlap/\Delta + \rlap/q_1 + i \epsilon} \gamma^\mu +
\gamma^\mu \frac{i}{\rlap/k + (1-\alpha) \rlap/\Delta - \rlap/q_1 + i \epsilon} \gamma^\nu\right] \underline{M}(k) \right\} \nonumber
\end{eqnarray}

and therefore

\begin{eqnarray}
T^{\mu \nu} &=& - \int \frac{d k\cdot n \ d k\cdot \tilde{n} \ d k_\perp^2}{(2 \pi)^4} \int \frac{d\lambda \ dx}{2 \pi}e^{i \lambda (x - k\cdot n)}  \int d^4z \ e^{i k\cdot z} \nonumber \\
&& \tr\left\{ \left[ \gamma^\nu
\frac{i}{\rlap/k - \alpha \rlap/\Delta + \rlap/q_1 + i \epsilon} \gamma^\mu +\gamma^\mu \frac{i}{\rlap/k + (1-\alpha) \rlap/\Delta - \rlap/q_1 + i \epsilon} \gamma^\nu\right] \right. \nonumber \\
&\cdot& \left. \langle P'|
\overline{\psi} (-\alpha z) \psi ((1-\alpha)z) |P \rangle \right\}. \nonumber
\end{eqnarray}

Keeping the leading terms for the loop momenta

\begin{eqnarray}
k^\mu - \alpha \Delta^\mu + q_1^\mu &=& \tilde{n}^\mu \left(k\cdot n + 2\alpha \xi - 2\xi\right) + n^\mu \left(k\cdot \tilde{n} - {\alpha}\xi \overline{M}^2 + \frac{Q^2}{4 \xi} \right) \nonumber  \\
k^\mu +(1-\alpha) \Delta^\mu - q_1^\mu &=& \tilde{n}^\mu \left( k\cdot n -2(1-\alpha) \xi + 2\xi \right) + 
n^\mu \left( k\cdot \tilde{n} +2(1-\alpha) \xi \overline{M}^2 - \frac{Q^2}{4 \xi} \right) \nonumber
\end{eqnarray}
we thus obtain

\begin{eqnarray}
T^{\mu \nu} &=&  - \int \frac{d (k\cdot n) \ d (k\cdot\tilde{n}) \ d k_\perp^2}{(2 \pi)^4} \int \frac{d\lambda \ dx}{2 \pi}e^{i \lambda (x - k\cdot n)}  \int d^4z \ e^{i k\cdot z} \nonumber \\
&& \tr\left\{ \left[ \gamma^\nu \frac{\rlap/\tilde{n}}{2 \left(k\cdot\tilde{n}  - {\alpha} \xi \overline{M}^2 + 
\frac{Q^2}{4 \xi} \right)} \gamma^\mu +\gamma^\mu \frac{\rlap/\tilde{n}}{2 \left( k\cdot\tilde{n}  +(1-\alpha){\xi} 
\overline{M}^2 - \frac{Q^2}{4 \xi} \right)} \gamma^\nu \right. \right.\nonumber \\
&+& \left. \left.\gamma^\nu \frac{\rlap/n}{2 \left( k\cdot n + 2\alpha \xi - 2\xi
\right)} \gamma^\mu +
\gamma^\mu \frac{\rlap/n}{2 \left( k\cdot n -2(1-\alpha) \xi + 2\xi\right)} \gamma^\nu \right] \right. \nonumber \\
&\cdot& \left. \langle P'| 
\overline{\psi} (-\alpha z) \psi ((1-\alpha)z) |P \rangle \right\}. \nonumber 
\end{eqnarray}

We expand

\begin{eqnarray}
k\cdot z &=& (k\cdot n) \ (\tilde{n}\cdot z) + (k\cdot\tilde{n}) \ (n\cdot z) - \vec{k}_\perp\cdot \vec{z}_\perp \nonumber
\end{eqnarray}

and choose $\alpha = 1/2$. These expansions are introduced in eq.~\ref{coll} and after some manipulations the tensor $T$ now becomes

\begin{eqnarray}
T^{\mu \nu} |_{\alpha=1/2} &=&  - 
\int \frac{d (k\cdot n) \ d (k\cdot\tilde{n} ) \ d k_\perp^2}{(2 \pi)^4} \int \frac{d\lambda \ dx}{2 \pi}e^{i \lambda (x - k\cdot n)}  \int d^4z \ e^{i k\cdot z} \nonumber \\
&& \tr\left\{ \left[ \gamma^\nu\frac{\rlap/\tilde{n}}{2 \left(k\cdot\tilde{n} - \frac{\alpha}{2} \xi \overline{M}^2 + 
\frac{Q^2}{2 \xi} \right)} \gamma^\mu +\gamma^\mu \frac{\rlap/\tilde{n}}{2 \left( k\cdot \tilde{n} +(1-\alpha){\xi} 
\overline{M}^2 - \frac{Q^2}{4 \xi} \right)} \gamma^\nu \right. \right.\nonumber \\
&+& \left. \left.\gamma^\nu \frac{\rlap/n}{2 \left( k\cdot n + 2\alpha \xi - 2\xi
\right)} \gamma^\mu +\gamma^\mu \frac{\rlap/n}{2 \left( 
k\cdot n -2(1-\alpha) \xi + 2\xi\right)} \gamma^\nu 
\right] \right. \nonumber \\
&\cdot& \left. \langle P'| 
\overline{\psi} (-\alpha z) \psi ((1-\alpha)z) |P \rangle \right\}. \nonumber
\end{eqnarray}

We also recall that the expansion of the matrix element $\underline{M}(k)$ proceeds also in this case as in the electromagnetic case 

\begin{eqnarray}
\underline{M}_{ab}^{(i)}(k) &=& \int d^4y e^{i k\cdot y} \langle P'| 
\overline{\psi}_{a}^{(i)}(-\alpha y) \psi_{b}^{(i)}((1-\alpha)y) |P \rangle = 
A_1 \slash{\tilde{n}} + A_2\gamma_5 \slash{\tilde{n}} +...  
\label{matrix}
\end{eqnarray}

where the ellipses refer to terms which are of higher twist or disappear in the trace of the diagram.

\section*{Appendix B}

The last ingredients needed in the construction of the input distribution functions are the form factors $F^i_1$ and $F^i_2$. From experimental measurements we know, by a dipole parametrization in the small $\Delta^2$ region, that

\begin{equation}
G_E^p (\Delta^2)= (1 + \kappa_p)^{-1} G_M^p (\Delta^2)= \kappa_n^{-1} G_M^n (\Delta^2)
= \left( 1 - \frac{\Delta^2}{m_V^2} \right)^{- 2} ,\qquad
G_E^n (\Delta^2) = 0,
\end{equation}

where the electric, $G_E^i (\Delta^2) = F_1^i (\Delta^2) + \frac{\Delta^2}{4 M^2} F_2^i (\Delta^2)$, and magnetic 
form factors $G_M^i (\Delta^2) = F_1^i (\Delta^2) + F_2^i (\Delta^2)$ are usually parametrized in terms 
of a cutoff mass $m_V = 0.84\, {\rm GeV}$. \newline 
For non-polarized GPD's the valence $u$ and $d$ quark form factors in the proton can be easily extracted from 
$F_I^{({p \atop n})} = 2 \left( {Q_u\atop Q_d} \right) F_I^u + \left( {Q_d \atop Q_u} \right) F_I^d$  and 
given exclicitely by 

\begin{equation}
2 F^u_I (\Delta^2) = 2 F_I^p (\Delta^2) + F_I^n (\Delta^2),
\qquad
F^d_I (\Delta^2) = F_I^p (\Delta^2) + 2 F_I^n (\Delta^2) ,
\quad\mbox{for}\quad I = 1,2.
\end{equation}

This exploits the fact that proton and neutron form an iso-spin doublet.

At the scale $m_{A}=0.9$ GeV one can get 

\ba
G^{i}_{1}(\Delta^2)=\left(1-\frac{\Delta^2}{m_{A}^{2}}\right)^{-2}
\ea

for the valence quarks.
For the form factors $F$ we obtain

\ba
&&F_1^u=-\frac{A \Delta^2}{M^2 (1 - B \Delta^2)^2 \left(-1 + \frac{\Delta^2}{4 M^2}\right)} + \frac{1 - \frac{C \Delta^2}{M^2\left(1 - \frac{\Delta^2}{4 M^2}\right)}}{(1 - B \Delta^2)^2},\nonumber\\
&&F_2^u=\frac{D}{(1 - B \Delta^2)^2\left(1 - \frac{\Delta^2}{4 M^2}\right)},\nonumber\\
&&F_1^d=-\frac{E \Delta^2}{M^2 (1 - B \Delta^2)^2\left(-1 + \frac{\Delta^2}{4 M^2}\right)}+\frac{1 - \frac{C \Delta^2}{M^2\left(1 - \frac{\Delta^2}{4 M^2}\right)}}{(1 - B \Delta^2)^2},\nonumber\\
&&F_2^d=\frac{F}{(1 - B \Delta^2)^2\left(1 - \frac{\Delta^2}{4 M^2}\right)},\nonumber\\
\ea

where

\ba
A=0.238\hspace{1cm} B=1.417\hspace{1cm} C=0.447,\nonumber\\ 
D=0.835\hspace{1cm} E=0.477\hspace{1cm} F=0.120.
\ea

\section*{Appendix C}

In this section we illustrate an analytical computation of the integrals by discretization, using finite elements method. 
We want to approximate with high numerical accuracy integrals of the form
\ba
P.V.\int_{-1}^{1}\frac{H(z)dz}{z-\xi}=\int_{-1}^{\xi}dz\frac{H(z)-H(\xi)}{z-\xi} +\int_{\xi}^{1}dz\frac{H(z)-H(\xi)}{z-\xi} + H(\xi)\ln\left|\frac{\xi-1}{\xi+1}\right|.
\ea

For this purpose we start by choosing a grid on the interval  $(-1=x_0,.....,x_{n+1}=\xi)$ and define

\ba
J_1=\int_{-1}^{\xi}dz\frac{H(z)-H(\xi)}{z-\xi}=\sum_{j=0}^{n}\int_{x_j}^{x_{j+1}}dx\frac{H(x)-H(\xi)}{x-\xi}\,.
\ea

Performing a simple linear interpolation we get 

\ba
J_1&=&\sum_{j=0}^{n-1}\int_{x_j}^{x_{j+1}} \left\{H(x_j)\left[\frac{x_{j+1}-x}{x_{j+1}-x_j}\right]+ H(x_{j+1})\left[\frac{x-x_{j}}{x_{j+1}-x_j}\right]\right\} \frac{dx}{x-\xi}\,\,\nonumber\\ 
&+&\int_{x_n}^{\xi}\left\{H(x_n)\left[\frac{\xi-x}{\xi-x_n}\right]+ H(\xi)\left[\frac{x-x_{n}}{\xi-x_n}\right]\right\} \frac{dx}{x-\xi} -\int_{-1}^{\xi}dx\frac{H(\xi)}{x-\xi}\,\,.
\ea
 
After the integration we are left with

\ba
J_1 &=& \sum_{j=0}^{n-1} H(x_j)\left[-1 +\left(\frac{x_{j+1}-\xi}{x_{j+1}-x_j}\right)\ln\left|\frac{x_{j+1}-\xi}{x_j-\xi}\right|\right]\nonumber\\
&+&\sum_{j=0}^{n-1}H(x_{j+1})\left[1 +\left(\frac{\xi-x_{j}}{x_{j+1}-x_j}\right)\ln\left|\frac{x_{j+1}-\xi}{x_j-\xi}\right|\right]\nonumber\\
&-& \sum_{j=0}^{n-1}H(\xi)\ln\left|\frac{x_{j+1}-\xi}{x_j-\xi}\right|-H(x_n)+H(\xi)\,\,. 
\ea 

Now, moving to the integral in the interval $(\xi,1)$, we introduce a similar grid of equally spaced points 
$(\xi=y_0,......,y_{n+1}=1)$ and define the integral

\ba
J_2=\int_{\xi}^{1}dz\frac{H(z)-H(\xi)}{z-\xi}=\sum_{j=0}^{n}\int_{y_j}^{y_{j+1}}dy\frac{H(y)-H(\xi)}{y-\xi}.
\ea

As above, after isolating the singularity we obtain 

\ba  
J_2&=&\sum_{j=1}^{n}\int_{y_j}^{y_{j+1}} \left\{H(y_j)\left[1-\frac{y-y_{j}}{y_{j+1}-y_j}\right]+ H(y_{j+1})\left[\frac{y-y_{j}}{y_{j+1}-y_j}\right]\right\} \frac{dy}{y-\xi}\,\,\nonumber\\ 
&+&H(y_1) + H(\xi)\int_{\xi}^{y_1}\left[\frac{y_1-y}{y_1-\xi}\right]dy\,\,-\,\,\int_{\xi}^{1}dy\frac{H(\xi)}{y-\xi}\,\,.
\ea

Again, performing the integrations we obtain

\ba 
J_2 &=& \sum_{j=1}^{n} H(y_j)\left[-1 +\left(\frac{y_{j+1}-\xi}{y_{j+1}-y_j}\right)\ln\left|\frac{y_{j+1}-\xi}{y_j-\xi}\right|\right]\nonumber\\
&+&\sum_{j=1}^{n}H(y_{j+1})\left[1 +\left(\frac{\xi-y_{j}}{y_{j+1}-y_j}\right)\ln\left|\frac{y_{j+1}-\xi}{y_j-\xi}\right|\right]\nonumber\\
&-&\sum_{j=1}^{n}H(\xi)\ln\left|\frac{y_{j+1}-\xi}{y_j-\xi}\right|+H(y_1)-H(\xi)\,\,. 
\ea

Collecting our results, at the end we obtain

\ba
P.V.\int_{-1}^{1}\frac{H(z)dz}{z-\xi}= J_1 + J_2 +H(\xi)\ln\left|\frac{\xi-1}{\xi+1}\right|\,.
\ea

We can use the same strategy for the integrals of ``$+$'' type defined as follows

\ba
P.V.\int_{-1}^{1}\frac{H(z)dz}{z+\xi}=\int_{-1}^{-\xi}dz\frac{H(z)-H(-\xi)}{z+\xi} +\int_{-\xi}^{1}dz\frac{H(z)-H(-\xi)}{z+\xi} + H(-\xi)\ln\left|\frac{\xi+1}{\xi-1}\right|.
\ea

This time we call our final integrals  $X_1$ and $X_2$. They are given by the expressions  

\ba
X_1&=&\sum_{j=0}^{n-1} H(x_j)\left[-1 +\left(\frac{x_{j+1}+\xi}{x_{j+1}-x_j}\right)\ln\left|\frac{x_{j+1}+\xi}{x_j+\xi}\right|\right]\nonumber\\
&+&\sum_{j=0}^{n-1}H(x_{j+1})\left[1 +\left(\frac{-\xi-x_{j}}{x_{j+1}-x_j}\right)\ln\left|\frac{x_{j+1}+\xi}{x_j+\xi}\right|\right]\nonumber\\
&-& \sum_{j=0}^{n-1}H(-\xi)\ln\left|\frac{x_{j+1}+\xi}{x_j+\xi}\right|-H(x_n)+H(-\xi)\,\,,
\ea

with a discretization supported in the $(-1=x_0,.....,x_{n+1}=\xi)$ grid, and 

\ba
X_2&=&\sum_{j=1}^{n} H(y_j)\left[-1 +\left(\frac{y_{j+1}+\xi}{y_{j+1}-y_j}\right)\ln\left|\frac{y_{j+1}+\xi}{y_j+\xi}\right|\right]\nonumber\\
&+&\sum_{j=1}^{n}H(y_{j+1})\left[1 +\left(\frac{-\xi-y_{j}}{y_{j+1}-y_j}\right)\ln\left|\frac{y_{j+1}+\xi}{y_j+\xi}\right|\right]\nonumber\\
&-&\sum_{j=1}^{n}H(-\xi)\ln\left|\frac{y_{j+1}+\xi}{y_j+\xi}\right|+H(y_1)-H(-\xi)\,\,.
\ea

on the $(-\xi=y_0,.....,y_{n+1}=1)$ grid. As a final result for the ``$+$'' integral we get 

\ba
P.V.\int_{-1}^{1}\frac{H(z)dz}{z+\xi}= X_1 + X_2 +H(-\xi)\ln\left|\frac{\xi+1}{\xi-1}\right|\,.
\ea

\section*{Appendix D}

In this section we will present the full expression of the functions $A_1$ and $A_2$ which appear in the squared amplitude 

\ba
A_1(z,z'x,t,Q^2)&=&\tilde{g}^{4} Q^{2}\left.\left[4 g_{d}^{2}[\tilde{E}_{d}' (4 \tilde{H}_{d}M^{2} + \tilde{E}_{d}t)x^{2} \right.\right.\nonumber\\ 
&&+\left.\left.4 \tilde{H}_{d}'M^{2} (4\tilde{H}_{d}(x - 1)+ \tilde{E}_{d}x^{2})] \right.\right.\nonumber \\
&&+\left.\left.4g_{d} g_{u}[(4 \tilde{E}_{u}' \tilde{H}_{d} M^{2} + 4 \tilde{E}_{d}' \tilde{H}_{u}M^{2} +  \tilde{E}_{u}' \tilde{E}_{d} t + \tilde{E}_{d}' \tilde{E}_{u}t)x^{2} \right.\right.\nonumber \\
&&+\left.\left.4 \tilde{H}_{u}'M^{2}(4 \tilde{H}_{d}(x - 1) +  \tilde{E}_{d}x^{2}) + 4\tilde{H}_{d}'M^{2}(4 \tilde{H}_{u} (x- 1) + \tilde{E}_{u}x^{2})]\right.\right.\nonumber \\
&&+ \left.\left.D_{v} U_{v}g_{d}g_{u}[4 E_{u}' E_{d} t +4 E_{d}' E_{u} t - 4  E_{u}' E_{d}tx - 4  E_{d}' E_{u}tx + 4  E_{u}' E_{d}M^{2}x^{2}\right.\right.\nonumber \\
&&+ \left.\left.4  E_{d}' E_{u} M^{2}x^{2}  + 4 E_{u}' H_{d}M^{2}x^{2} + 4 E_{d}' H_{u}M^{2}x^{2} + E_{u}' E_{d}tx^{2} +E_{d}'E_u t x^2 \right.\right.\nonumber \\
&&+ \left.\left.4 H_{u}'M^{2}(4 H_{d}(x - 1) + E_{d}x^{2}) + 4 H_{d}'M^{2}(4 H_{u} (x - 1)+E_{u}x^{2})] \right.\right.\nonumber \\
&&+\left.\left.g_{u}^{2}[4E_{u}'E_{u}t U_{v}^{2} - 4 E_{u}'E_{u} t U_{v}^{2}x +16 \tilde{E}_{u}'\tilde{H}_{u}M^{2}x^{2} + 4\tilde{E}_{u}'\tilde{E}_{u} t x^{2} \right.\right.\nonumber \\
&&+\left.\left.4  E_{u}' E_{u}M^{2} U_{v}^{2} x^{2} + 4 E_{u}' H_{u}M^{2} U_{v}^{2} x^{2} +E_{u}'E_{u} t U_{v}^{2} x^{2} + 16\tilde{H}_{u}'M^{2}(4\tilde{H}_{u}(x - 1) + \tilde{E}_{u}x^{2})\right.\right.\nonumber \\
&&+ \left.\left.4H_{u}'M^{2} U_{v}^{2}(4H_{u}(x - 1) + E_{u}x^{2})] \right.\right.\nonumber\\
&&+\left.\left.D_{v}^{2} g_{d}^{2} [4H_{d}' M^{2}(4H_{d}(x - 1) + E_{d}x^{2}) + E_{d}'(4H_{d}M^{2}x^{2}+ E_{d}(t \,(x -2)^{2} + 4M^{2} x^{2}))]\right]\right.\nonumber\\
\ea

and for $A_2(z,z',x,t)$ we get a similar result

\ba
A_2(z,z',x,t,Q^2) &=&\left.4\tilde{g}^4 Q^2\left[g_d  g_u [4 E'_{u} E_d  t + 4 E'_{d} E_u  t - 16 D_v \tilde{H}'_u \tilde{H}_d  M^2  U_v - 16  D_v \tilde{H}'_d  \tilde{H}_u  M^2 U_v - 4 E'_u E_d t x \right.\right.\nonumber\\
&-&\left.\left.4 E'_d E_u t x + 16 D_v \tilde{H}_u \tilde{H}_d M^2 U_v x + 16 D_v \tilde{H}'_d \tilde{H}_u M^2 U_v x + 4 E'_u E_d M^2 x^2 + 4 E'_d E_u M^2 x^2 \right. \right.\nonumber\\
&+&\left.\left.4 E'_u H_d M^2 x^2 + 4 E'_d H_u M^2 x^2 + E'_u E_d t x^2 + E'_d E_u t x^2 + 4 D_v \tilde{E}_u \tilde{H}'_d M^2 U_v x^2 \right.\right.\nonumber\\
&+&\left.\left.4 D_v \tilde{E}_d \tilde{H}'_u M^2 U_v x^2 + 4 D_v \tilde{E}'_u \tilde{H}_d M^2 U_v x^2 +
4 D_v \tilde{E}'_d \tilde{H}_u M^2 U_v x^2 + D_v \tilde{E}'_u \tilde{E}_d t U_v x^2 \right.\right.\nonumber\\
&+&\left.\left.D_v \tilde{E}'_d \tilde{E}_u t U_v x^2 + 4 H'_u M^2 (4 H_d ( x-1) + E_d x^2) + 4 H'_d M^2 (4 H_u (x-1)  + E_u x^2)]\right.\right.\nonumber\\
&+&\left.\left.g_d^2 [4H'_d M^2 (4 H_d(x-1) + E_d x^2) + D_{v}^{2}(\tilde{E}'_d (4\tilde{H}_d M^2 + \tilde{E}_d t) x^2 \right.\right.\nonumber\\
&+&\left.\left.4\tilde{H}'_d M^2 (4 \tilde{H}_d (x-1) + \tilde{E}_d x^2)) + E'_d (4 H_d M^2 x^2 + E_d(t (x-2)^2 + 4 M^2 x^2))]\right.\right.\nonumber\\
&+&\left.\left.g_{u}^{2}[4 H'_u M^2 (4 H_u (x-1) + E_u x^2) + U_{v}^{2} (\tilde{E}'_u (4 \tilde{H}_u M^2 + \tilde{E}_u t) x^2 \right.\right.\nonumber\\
&+&\left.\left.4 \tilde{H}'_{u} M^2 (4 \tilde{H}_u(x-1) + \tilde{E}_u x^2)) + E'_{u} (4 H_u M^2 x^2 + E_u (t(x-2)^2 + 4 M^2 x^2))] \right] \right..\nonumber\\
\ea

\end{document}